\documentclass[aps,twocolumn,twoside,preprintnumbers,nofootinbib,superscriptaddress]{revtex4}

\usepackage{amsmath}
\usepackage{amssymb}
\usepackage{amsbsy}
\usepackage{mathptmx}
\usepackage{graphicx}

\numberwithin{equation}{section}

\begin{document}
\preprint{KOBE-TH-07-06}
\title{Supersymmetry in 5d Gravity}
\author{C. S. Lim}
\email{lim@kobe-u.ac.jp}
\affiliation{Department of Physics, Kobe University, 1-1 Rokkodai, Nada, Kobe 657-8501, Japan}
\author{Tomoaki Nagasawa}
\email{nagasawa@anan-nct.ac.jp}
\affiliation{Anan National College of Technology, 265 Aoki, Minobayashi, Anan 774-0017, Japan}
\author{Satoshi Ohya}
\email{ohya@kobe-u.ac.jp}
\affiliation{Graduate School of Science, Kobe University, 1-1 Rokkodai, Nada, Kobe 657-8501, Japan}
\author{Kazuki Sakamoto}
\email{049d841n@stu.kobe-u.ac.jp}
\affiliation{Graduate School of Science, Kobe University, 1-1 Rokkodai, Nada, Kobe 657-8501, Japan}
\author{Makoto Sakamoto}
\email{dragon@kobe-u.ac.jp}
\affiliation{Department of Physics, Kobe University, 1-1 Rokkodai, Nada, Kobe 657-8501, Japan}
\date{October 2, 2007}
\begin{abstract}
We study a 5d gravity theory with a warped metric and show that two $N = 2$ supersymmetric quantum-mechanical systems are hidden in the 4d spectrum.
The supersymmetry can be regarded as a remnant of higher-dimensional general coordinate invariance and turns out to become a powerful tool to determine the physical 4d spectrum and the allowed boundary conditions.
Possible extensions of the $N = 2$ supersymmetry are briefly discussed.
\end{abstract}
\maketitle
\section{INTRODUCTION} \label{sec:intro}
Over the past decade a considerable number of studies have been made on gauge/gravity theories with extra dimensions.
In gauge-Higgs unification scenario, extra components of gauge fields play a role of Higgs fields \cite{Hosotani:1983,Hosotani:1989,HIL:1998,ABQ:2001,KLY:2002,HNS:2002,BN:2003,HHKY:2004,HTY:2005,HNT:2004,HNT:2005}.
Attractive models of grand unified theories (GUTs) on orbifolds have been constructed, avoiding common problems of four-dimensional GUTs \cite{Kawamura1:2001,Kawamura2:2001,HN:2001,AF:2001,HM-R:2001}.
Higgsless gauge symmetry breaking can be realized via boundary conditions of extra dimensions \cite{NS:2004,Csaki:2003dt,Chivukula:2004pk,Georgi:2004iy,Foadi:2003xa,Csaki:2003zu,Nomura:2003du,Gabriel:2004ua,Schwinn:2004xa,Cacciapaglia:2004jz,Burdman:2003ya,Csaki:2004sz} and a large mass hierarchy can naturally be obtained in this scenario \cite{NS:2004}.
A higher dimensional scenario with a warped geometry has been proposed to solve the hierarchy problem by Randall-Sundrum \cite{RS1:1999}.
In the scenario, all scales except for the scale of gravity are reduced to the weak scale by a warped factor.
According to this scenario, various attempts have been made to construct realistic models \cite{Csaki:2003zu,Nomura:2003du,Goldberger:2002pc,Pomarol:1999ad,Davoudiasl:1999tf}.
Randall and Sundrum have also 
proposed a mechanism to localize gravity in the vicinity of a brane \cite{RS2:1999} which has attracted enormous attention
\cite{Kim:1999ja,Lykken:1999nb,Goldberger:1999uk,Chang:1999nh,Davoudiasl:1999tf,DeWolfe:1999cp,Charmousis:1999rg,Csaki:1999jh}.

In constructing realistic models with extra dimensions, the spectrum of light Kaluza-Klein (KK) modes becomes important if those masses are accessible to future collider experiments.
Thus, it will be worthwhile investigating what are characteristic features of the 4d spectrum of KK modes coming from extra dimensions.
A related subject to study is to clarify the cancellation mechanism of divergences in loop corrections.
Mass corrections to extra components of gauge fields are found to be finite at least at one-loop order.
The finiteness is very important because finite quantities can be considered to be predictions of higher-dimensional theories though they will not be renormalizable.
The cancellation of would-be divergences has not been, however, understood fully yet.
\footnote{Some development has been given in \cite{MY:2006,Hosotani:2006,Hosotani:2007kn}.}
In a 4-dimensional point of view, the cancellation of divergences seems to be mysterious because it occurs only after {\em all} massive KK modes are taken into account.
If we truncate massive KK modes at some energy, the cancellation becomes incomplete.
Furthermore, the cancellation still occurs even when gauge symmetries are broken via orbifolding, the Hosotani mechanism, or boundary conditions.
Therefore, it would be of great importance to reveal nontrivial structure hidden in the spectrum of KK modes and interactions between them.

A secret of gauge theories with extra dimensions has been uncovered in \cite{LNSS:2005}.
\footnote{Howe et al. \cite{HPPT:1989} discussed an $N = 2$ worldline supersymmetry for a realistic spin $N/2$ particle and succeeded to present field equations for massless and massive antisymmetric tensors in arbitrary space-time dimensions.
The $N = 2$ worldline supersymmetry seems to have some connections to the $N = 2$ supersymmetry found in \cite{LNSS:2005}, but a direct relation between them is not clear.}
It has been shown that an $N = 2$ supersymmetric quantum-mechanical system \cite{Witten:1981} is hidden in the 4d spectrum of any gauge invariant theory with extra dimensions.
The $N = 2$ supersymmetry can be regarded as a remnant of the higher-dimensional gauge invariance.
Our purpose of this paper is to extend the analysis of \cite{LNSS:2005} to the 5d gravity theory with the Randall-Sundrum metric and show that {\em two} $N = 2$ supersymmetric quantum-mechanical systems are hidden in the 4d spectrum of the model.
A part of the supersymmetric structure has already been pointed out in the literature \cite{DeWolfe:1999cp,Miemiec:2001}.
Those authors have noticed that the Hamiltonian of the mass eigenfunctions for massless/massive 4d gravitons can be written in a supersymmetric form $H = D^{\dagger}D$.
They have not, however, found its superpartner in the system and also missed another $N = 2$ supersymmetric system.
The authors have used supersymmetry mainly as a technical tool to solve the eigenvalue equations, especially the zero modes.
In this paper, we show that the mass eigenfunctions for the metric fluctuation fields are governed by two quantum-mechanical systems with full $N = 2$ supersymmetry, and further show that the supersymmetry can be a powerful tool to determine the 4d spectrum and the allowed boundary conditions.

This paper is organized as follows.
We consider a 5d pure Abelian gauge theory with a warped metric in Section \ref{sec:5dGauge}.
The results are not new but the purpose of this section is to show differences as well as resemblances between the 5d gauge theory and the 5d gravity one clearly.
In Section \ref{sec:5dGravity}, we investigate the 5d gravity theory with the Randall-Sundrum metric and discuss the supersymmetric structure, allowed boundary conditions compatible with the supersymmetry, and the physical spectrum in detail.
Section \ref{sec:concl} is devoted to conclusions and discussions.
\section{5d PURE ABELIAN GAUGE THEORY} \label{sec:5dGauge}
In \cite{LNSS:2005}, it has been shown that any gauge invariant theory with extra dimensions possesses a quantum-mechanical supersymmetric structure in the spectrum of the KK modes.
Following the analysis given in \cite{LNSS:2005}, we consider the pure Abelian gauge theory with a single extra dimension.
All the results in this section are not new but the purpose to derive them is to make differences as well as resemblances clear between the 5d gauge theory and the 5d gravity theory discussed in the next section.

Let us consider the 5d pure Abelian gauge theory with a single extra dimension compactified on an interval
\begin{align}
S
= 	\int{\rm d}^{4}x \int_{z_{1}}^{z_{2}}{\rm d}z\sqrt{-G}
	\left\{- \frac{1}{4}G^{MN}G^{KL}F_{MK}F_{NL}\right\}. \label{eq:gauge01}
\end{align}
We choose the background metric as
\footnote{$\eta_{\mu\nu} = {\rm diag}(-1, 1, 1, 1)$.}
\begin{align}
{\rm d}s^{2}
= 	{\rm e}^{2A(z)}(\eta_{\mu\nu}{\rm d}x^{\mu}{\rm d}x^{\nu} + {\rm d}z^{2}) \quad
\text{with}	 \quad
A(z) = -\ln\left(\frac{z}{z_{1}}\right). \label{eq:gauge02}
\end{align}
The metric describes the warped geometry in the conformal coordinate discussed by Randall-Sundrum \cite{RS1:1999,RS2:1999} and will also be used in the next section.
The $x^{\mu}$ ($\mu = 0, 1, 2, 3$) are the 4-dimensional Minkowski coordinates and $z$ is the extra dimensional one.
The extra dimension has two boundaries at $z = z_{1}$ and $z_{2}$.
The boundary conditions for the gauge fields will be determined later.

It follows from Eq.\eqref{eq:gauge02} that the 5d metric $G_{MN}$ has the form
\begin{align}
G_{MN}
= 	\begin{pmatrix}
	G_{\mu\nu} 	& G_{\mu5} \\
	G_{5\nu} 		& G_{55}
	\end{pmatrix}
= 	\begin{pmatrix}
	{\rm e}^{2A}\eta_{\mu\nu} 	& 0 \\
	0 						& {\rm e}^{2A}
	\end{pmatrix}.
\end{align}
Then, the action \eqref{eq:gauge01} reduces to
\begin{align}
S
&= 	\int{\rm d}^{4}x \int_{z_{1}}^{z_{2}}\!\!\!{\rm d}z~{\rm e}^{A}
	\left\{- \frac{1}{4}\eta^{\mu\nu}\eta^{\rho\sigma}F_{\mu\rho}F_{\nu\sigma}
	- \frac{1}{2}\eta^{\mu\nu}F_{\mu5}F_{\nu5}\right\} \nonumber\\
&= 	\int{\rm d}^{4}x \int_{z_{1}}^{z_{2}}\!\!\!{\rm d}z~{\rm e}^{A}
	\biggl\{
	\frac{1}{2}A_{\mu}
	\left[\eta^{\mu\nu}\bigl(\Box + (\partial_{z} + A')\partial_{z}\bigr) - \partial^{\mu}\partial^{\nu}\right]
	A_{\nu} \nonumber\\
& 	\hspace{1em}
	- \frac{1}{2}A_{\mu}(\partial_{z} + A')\partial^{\mu}A_{5}
	- \frac{1}{2}A_{5}\partial^{\mu}\partial_{z}A_{\mu}
	+ \frac{1}{2}A_{5}\Box A_{5}\biggr\}, \label{eq:gauge04}
\end{align}
where $\Box = \partial_{\mu}\partial^{\mu}$ and at the second equality we have integrated by parts and ignored boundary terms.
To obtain the 4d spectrum, we expand $A_{\mu}(x,z)$ and $A_{5}(x,z)$ as
\begin{align}
A_{\mu}(x,z)
&= 	\sum_{n}A_{\mu}^{(n)}(x)f^{(n)}(z), \label{eq:gauge05}\\
A_{5}(x,z)
&= 	\sum_{n}A_{5}^{(n)}(x)g^{(n)}(z). \label{eq:gauge06}
\end{align}
The mode functions $f^{(n)}$ and $g^{(n)}$ are taken to be the eigenfunctions of the Schr\"odinger-like equations
\begin{align}
D^{\dagger}Df^{(n)}(z)
&= 	m_{n}^{2}f^{(n)}(z), \label{eq:gauge07}\\
DD^{\dagger}g^{(n)}(z)
&= 	m_{n}^{2}g^{(n)}(z), \label{eq:gauge08}
\end{align}
where
\footnote{We will use the notation; $A^{\prime}(z) = \frac{\rm d}{{\rm d}z}A(z)$, $A^{\prime\prime}(z) = \frac{{\rm d}^{2}}{{\rm d}z^{2}}A(z)$, etc.}
\begin{align}
D = \partial_{z}, \quad
D^{\dagger} = -\bigl(\partial_{z} + A'(z)\bigr). \label{eq:gauge09}
\end{align}
Although the above notation for $D$ and $D^{\dagger}$ seems to be strange, 
$D^{\dagger} = -(\partial_{z} + A')$ is actually hermitian conjugate to 
$D = \partial_{z}$ with respect to the inner product
\footnote{Precisely speaking, to justify the statement we have to specify the boundary conditions which assure that no boundary terms appear in integration by parts.
This will be verified later.}
\begin{align}
\langle \psi | \varphi \rangle
= 	\int_{z_{1}}^{z_{2}}\!\!\!{\rm d}z~{\rm e}^{A}\psi(z)^{*}\varphi(z). \label{eq:gauge10}
\end{align}
We note that the factor ${\rm e}^{A}$ comes from the expression of the action \eqref{eq:gauge04}.
Therefore, as was shown in \cite{LNSS:2005}, the equations \eqref{eq:gauge07} and \eqref{eq:gauge08} can be unified into a supersymmetric form
\begin{align}
H\Psi^{(n)}(z) = m_{n}^{2}\Psi^{(n)}(z), \label{eq:gauge11}
\end{align}
where $H$ is the Hamiltonian
\begin{align}
H
= 	\begin{pmatrix}
	D^{\dagger}D 	& 0 \\
	0 			& DD^{\dagger}
	\end{pmatrix}
= 	\{Q, Q^{\dagger}\} \label{eq:gauge12}
\end{align}
and the supercharges $Q$, $Q^{\dagger}$ are defined by
\begin{align}
Q
= 	\begin{pmatrix}
	0 	& 0 \\
	D 	& 0
	\end{pmatrix}, \quad
Q^{\dagger}
= 	\begin{pmatrix}
	0 	& D^{\dagger} \\
	0 	& 0
	\end{pmatrix}. \label{eq:gauge13}
\end{align}
These operators act on two-component vectors
\begin{align}
\Psi(z)
= 	\begin{pmatrix}
	f(z) \\
	g(z)
	\end{pmatrix}
\end{align}
with the inner product
\begin{align}
\langle \Psi_{1} | \Psi_{2} \rangle
= 	\int_{z_{1}}^{z_{2}}\!\!\!{\rm d}z~{\rm e}^{A}
	\bigl\{f_{1}^{*}(z)f_{2}(z) + g_{1}^{*}(z)g_{2}(z)\bigr\}. \label{eq:gauge15}
\end{align}
It follows that the eigenvalue $m_{n}^{2}$ for $f^{(n)}$ and $g^{(n)}$ is doubly degenerate except for $m_{0} = 0$ and that they are related each other as
\begin{align}
Df^{(n)}(z) &= m_{n}g^{(n)}(z), \nonumber\\
D^{\dagger}g^{(n)}(z) &= m_{n}f^{(n)}(z), \label{eq:gauge16}
\end{align}
or equivalently,
\begin{align}
Q
\begin{pmatrix}
f^{(n)}(z) \\
0
\end{pmatrix}
&= 	m_{n}
	\begin{pmatrix}
	0 \\
	g^{(n)}(z)
	\end{pmatrix}, \nonumber\\
Q^{\dagger}
\begin{pmatrix}
0 \\
g^{(n)}(z)
\end{pmatrix}
&= 	m_{n}
	\begin{pmatrix}
	f^{(n)}(z) \\
	0
	\end{pmatrix} \label{eq:gauge17}
\end{align}
with appropriate normalizations.

It should be emphasized that the existence/nonexistence of the zero mode depends on the boundary conditions at $z = z_{1}$, $z_{2}$.
It follows from the analysis in \cite{LNSS:2005} \cite{NST:2003,NST:2004,NST:2005} that only the following types of boundary conditions are compatible with the supersymmetry:
\begin{align}
\text{Type (N, N) :}
&	\begin{cases}
	\partial_{z}f(z_{1}) = \partial_{z}f(z_{2}) = 0, \\
	g(z_{1}) = g(z_{2}) = 0,
	\end{cases} \label{eq:gauge18}\\
\text{Type (D, D) :}
&	\begin{cases}
	f(z_{1}) = f(z_{2}) = 0, \\
	(\partial_{z} + A')g(z_{1}) = (\partial_{z} + A')g(z_{2}) = 0,
	\end{cases} \label{eq:gauge19}\\
\text{Type (N, D) :}
&	\begin{cases}
	\partial_{z}f(z_{1}) = f(z_{2}) = 0, \\
	g(z_{1}) = (\partial_{z} + A')g(z_{2}) = 0,
	\end{cases} \label{eq:gauge20}\\
\text{Type (D, N) :}
&	\begin{cases}
	f(z_{1}) = \partial_{z}f(z_{2}) = 0, \\
	(\partial_{z} + A')g(z_{1}) = g(z_{2}) = 0.
	\end{cases} \label{eq:gauge21}
\end{align}
We then find that with the boundary conditions the Hamiltonian $H$ is hermitian and the supercharges $Q$, $Q^{\dagger}$ are hermitian conjugate to each other, as announced before.
We should note that $Q\Psi(z)$ and $Q^{\dagger}\Psi(z)$ satisfy the same boundary conditions as $\Psi(z)$, otherwise the supercharges would be ill-defined.

\begin{figure*}[t]
	\begin{center}
		\begin{minipage}{.24\hsize}
			\begin{center}
			\includegraphics[scale=.6]{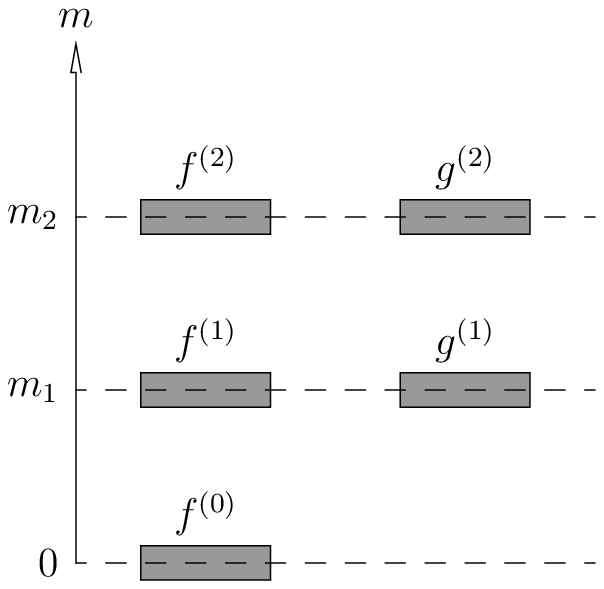}
			\vskip 1.5ex
			(a) Type (N, N)
			\end{center}
		\end{minipage}
		\begin{minipage}{.24\hsize}
			\begin{center}
			\includegraphics[scale=.6]{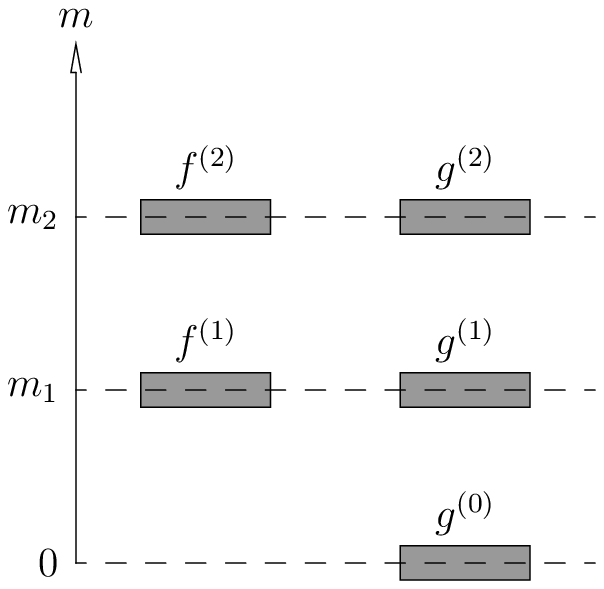}
			\vskip 1.5ex
			(b) Type (D, D)
			\end{center}
		\end{minipage}
		\begin{minipage}{.24\hsize}
			\begin{center}
			\includegraphics[scale=.6]{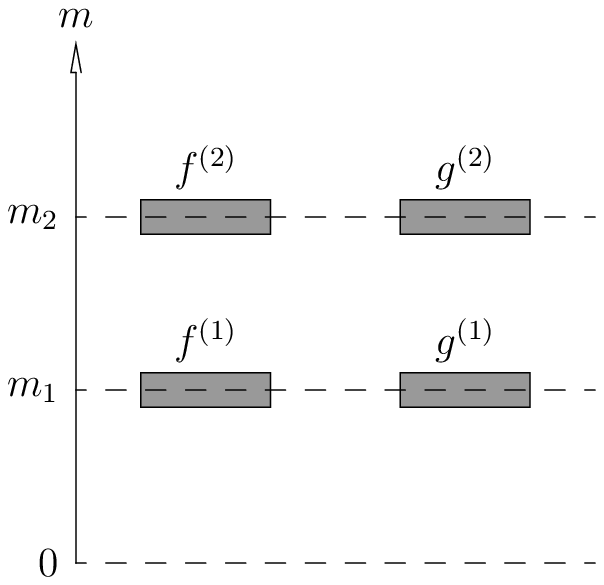}
			\vskip 1.5ex
			(c) Type (N, D)
			\end{center}
		\end{minipage}
		\begin{minipage}{.24\hsize}
			\begin{center}
			\includegraphics[scale=.6]{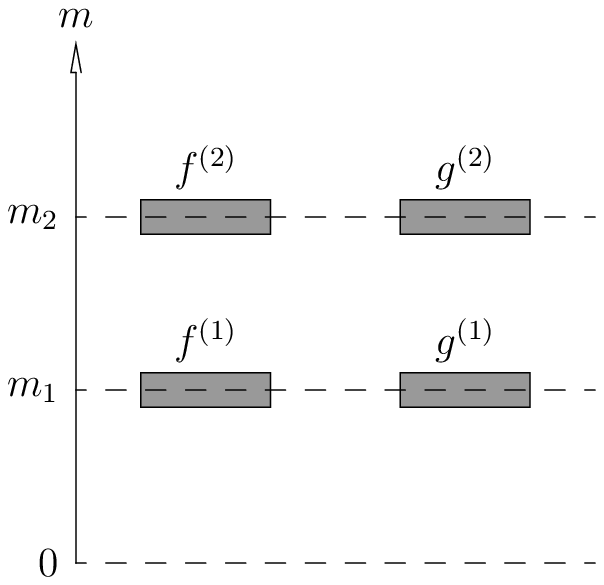}
			\vskip 1.5ex
			(d) Type (D, N)
			\end{center}
		\end{minipage}
	\caption
	{%
	A typical spectrum of $f^{(n)}$ and $g^{(n)}$.
	A zero mode $f^{(0)}$ ($g^{(0)}$) appears for the type (N, N) (type (D, D)) boundary conditions, while there is no zero mode for other boundary conditions.
	All nonzero modes are doubly degenerate between $f^{(n)}$ and $g^{(n)}$.
	}
	\label{fig:spectrum}
	\end{center}
\end{figure*}

The zero mode solutions with $m_{0} = 0$, if exists, should satisfy the first order differential equations
\begin{align}
& \partial_{z}f^{(0)}(z) = 0, \nonumber\\
& (\partial_{z} + A')g^{(0)}(z) = 0. \label{eq:gauge22}
\end{align}
The equations are easily solved as
\begin{align}
f^{(0)}(z) &= C^{(0)}, \nonumber\\
g^{(0)}(z) &= C^{\prime(0)}{\rm e}^{-A(z)}, \label{eq:gauge23}
\end{align}
where $C^{(0)}$ and $C^{\prime(0)}$ are normalization constants.
The solution $f^{(0)}$ ($g^{(0)}$) obeys only the type (N, N) (type (D, D)) boundary conditions.
Therefore, the zero mode with $m_{0} = 0$ exists for $f^{(0)}$ ($g^{(0)}$) with the type (N, N) (type (D, D)) boundary conditions, and there is no zero mode for other boundary conditions.
The results are summarized in FIG.\ref{fig:spectrum}.

Inserting the mode expansions \eqref{eq:gauge05}  and \eqref{eq:gauge06} into the action \eqref{eq:gauge04} and using the orthonormal relations of the mode functions with the relations \eqref{eq:gauge16}, we have
\begin{align}
S = \int{\rm d}^{4}x \bigl\{{\cal L}_{m = 0} + {\cal L}_{m\neq0}\bigr\}, \label{eq:gauge24}
\end{align}
where
\begin{align}
{\cal L}_{m=0}
&= 	\begin{cases}
	- \frac{1}{4}(\partial_{\mu}A_{\nu}^{(0)} - \partial_{\nu}A_{\mu}^{(0)})^{2} 	& \text{for Type (N, N)}, \\
	- \frac{1}{2}(\partial_{\mu}A_{5}^{(0)})^{2} 							& \text{for Type (D, D)}, \\
	0 															& \text{for Type (N ,D) or (D, N)}, 
	\end{cases} \label{eq:gauge25}\\
{\cal L}_{m\neq0}
&= 	\sum_{n=1}^{\infty}
	\Biggl\{
	- \frac{1}{4}(\partial_{\mu}A_{\nu}^{(n)} - \partial_{\nu}A_{\mu}^{(n)})^{2}
	- \frac{1}{2}m_{n}^{2}\left(A_{\mu}^{(n)} - \frac{1}{m_{n}}\partial_{\mu}A_{5}^{(n)}\right)^{2}
	\Biggr\}. \label{eq:gauge26}
\end{align}

We should make a few comments here.
The 4d gauge symmetry is broken except for the type (N, N) boundary conditions because there is no massless vector for other boundary conditions \cite{NS:2004,Csaki:2003dt,Chivukula:2004pk,Georgi:2004iy,Foadi:2003xa,Csaki:2003zu,Nomura:2003du,Gabriel:2004ua,Schwinn:2004xa,Cacciapaglia:2004jz,Burdman:2003ya,Csaki:2004sz}.
The modes $A_{5}^{(n)}$ ($n\neq0$) appear in the action only in the combinations $A_{\mu}^{(n)} - \frac{1}{m_{n}}\partial_{\mu}A_{5}^{(n)}$.
This implies that $A_{5}^{(n)}$ for $n\neq0$ are unphysical and can be absorbed into $A_{\mu}^{(n)}$ by gauge transformations, which are called the unitary gauge.
On the other hand, the zero mode $A_{5}^{(0)}$ for the type (D, D) boundary conditions cannot be removed from the action, so that it is a physical degree of freedom.
\section{5d GRAVITY with RANDALL-SUNDRUM BACKGROUND} \label{sec:5dGravity}
In this section, we investigate the supersymmetric structure of the 5d gravity theory with the Randall-Sundrum background metric \cite{RS1:1999,RS2:1999} in detail.
The analysis will make differences as well as resemblances clear between the 5d gravity theory and the 5d gauge theory discussed in the previous section.
\subsection{Set up} \label{sec:setup}
We consider a five-dimensional braneworld gravity with a single extra dimension compactified on an interval \cite{RS1:1999}
\footnote
{%
In this paper, we use the convention:
\begin{align}
\Gamma^{A}_{MN}
&= 	\frac{1}{2}G^{AB}(\partial_{N}G_{BM} + \partial_{M}G_{BN} - \partial_{B}G_{MN}), \nonumber\\
{R^{K}}_{LMN}
&= 	\partial_{M}\Gamma^{K}_{LN} - \partial_{N}\Gamma^{K}_{LM}
	+ \Gamma^{A}_{LN}\Gamma^{K}_{AM} - \Gamma^{A}_{LM}\Gamma^{K}_{NA}, \nonumber\\
R_{MN}
&= 	{R^{A}}_{MAN}. \nonumber
\end{align}
}:
\begin{align}
S
&= 	\int{\rm d}^{4}x \int_{z_{1}}^{z_{2}}\!\!\!{\rm d}z
	\sqrt{-G}(M^{3}R - \Lambda) \nonumber\\
&	\hspace{1em}
	+ \int{\rm d}^{4}x
	\sqrt{-g_{\rm UV}}(-\sigma_{\rm UV})\Bigr|_{z=z_{1}}
 	+ \int{\rm d}^{4}x 
	\sqrt{-g_{\rm IR}}(-\sigma_{\rm IR})\Bigr|_{z=z_{2}} , 
\label{eq:gravity01}
\end{align}
where $z_{i}$ ($i = 1, 2$) are the locations of the two branes, and $M$ is the five-dimensional Planck scale.
The $g^{\rm UV}_{\mu\nu}$ ($g^{\rm IR}_{\mu\nu}$) is the metric induced on the UV (IR) brane.
The bulk cosmological constant $\Lambda$ and the brane tensions $\sigma_{\rm UV}$ and $\sigma_{\rm IR}$ are tuned to give the warped metric
\begin{align}
{\rm d}s^{2}
= {\rm e}^{2A(z)}(\eta_{\mu\nu}{\rm d}x^{\mu}{\rm d}x^{\nu} + {\rm d}z^{2}), \label{eq:gravity02}
\end{align}
where
\begin{align}
A(z)
&= 	- \ln\left(\frac{z}{z_{1}}\right), \nonumber\\
\frac{1}{z_{1}}
&= 	\sqrt{\frac{-\Lambda}{12M^{3}}}, \nonumber\\
\sigma_{\rm UV}
&= 	- \sigma_{\rm IR} = \frac{12M^{3}}{z_{1}}. \label{eq:gravity03}
\end{align}
Here, the location of the UV brane is chosen such that  the warp factor is set to equal to $1$ on the UV brane $(z = z_{1})$.
\subsection{Quadratic Action} \label{sec:action}
We investigate the gravitational fluctuations around the RS background solution \eqref{eq:gravity02}:
\begin{align}
{\rm d}s^{2}
= 	{\rm e}^{2A}(\eta_{MN} + {\bar h}_{MN}){\rm d}x^{M}{\rm d}x^{N}. \label{eq:gravity04}
\end{align}
The action is invariant under infinitesimal general coordinate transformations
\begin{align}
x^{M} \to x^{M} + \xi^{M}(x) \label{eq:gravity05}
\end{align}
which are translated into the transformations of the metric fluctuations
\begin{align}
{\bar h}_{\mu\nu}
&\to 	{\bar h}_{\mu\nu}
	- \partial_{\mu}\xi_{\nu} - \partial_{\nu}\xi_{\mu} - 2A'\xi_{5}\eta_{\mu\nu}, \label{eq:gravity06}\\
{\bar h}_{\mu5}
&\to 	{\bar h}_{\mu5}
	- \partial_{z}\xi_{\mu} - \partial_{\mu}\xi_{5}, \label{eq:gravity07}\\
{\bar h}_{55}
&\to 	{\bar h}_{55}
	- 2(\partial_{z} + A')\xi_{5}. \label{eq:gravity08}
\end{align}
One might expand the fluctuation fields ${\bar h}_{MN}(x,z)$ as
\begin{align}
{\bar h}_{\mu\nu}(x,z)
&= 	\sum_{n}{\bar h}_{\mu\nu}^{(n)}(x){\bar f}^{(n)}(z), \nonumber\\
{\bar h}_{\mu5}(x,z)
&= 	\sum_{n}{\bar h}_{\mu5}^{(n)}(x){\bar g}^{(n)}(z), \nonumber\\
{\bar h}_{55}(x,z)
&= 	\sum_{n}{\bar h}_{55}^{(n)}(x){\bar k}^{(n)}(z) \nonumber
\end{align}
with some definite boundary condition for each mode function of ${\bar f}^{(n)}(z)$, ${\bar g}^{(n)}(z)$ and ${\bar k}^{(n)}(z)$.
It turns out, however, that the last term in Eq.\eqref{eq:gravity06} is incompatible with the above mode expansions.
The compatibility between the general coordinate transformations \eqref{eq:gravity06} -- \eqref{eq:gravity08} and the mode expansions leads to the following parameterization of ${\bar h}_{MN}$:
\begin{align}
{\bar h}_{MN}
= 	\begin{pmatrix}
	h_{\mu\nu} - \frac{1}{2}\eta_{\mu\nu}\phi 	& h_{\mu5} \\
	h_{5\nu} 								& \phi
	\end{pmatrix}. \label{eq:gravity09}
\end{align}
Then, the quadratic action of the metric fluctuations is found to be of the form
\begin{widetext}
\begin{align}
S^{(2)}
&= 	M^{3}\!\!\int{\rm d}^{4}x\int_{z_{1}}^{z_{2}}\!\!\!{\rm d}z~{\rm e}^{3A}
	\biggl\{\frac{1}{2}h_{\mu\nu}K^{\mu\nu; \rho\sigma}h_{\rho\sigma}
	+ 2h_{\mu5}K^{\mu5; \rho5}h_{\rho5}
	+ \frac{1}{2}\phi K^{\phi; \phi}\phi \nonumber\\
& 	\hspace{9em}
	+ h_{\mu\nu}K^{\mu\nu; \rho5}h_{\rho5}
	+ h_{\mu5}K^{\mu5; \rho\sigma}h_{\rho\sigma}
	+ \frac{1}{2}h_{\mu\nu}K^{\mu\nu; \phi}\phi \nonumber\\
& 	\hspace{9em}
	+ \frac{1}{2}\phi K^{\phi; \rho\sigma}h_{\rho\sigma}
	+ h_{\mu5}K^{\mu5; \phi}\phi
	+ \phi K^{\phi; \rho5}h_{\rho5}\biggr\}, \label{eq:gravity10}
\end{align}
where
\begin{align}
K^{\mu\nu; \rho\sigma}
&= 	- \frac{1}{4}
	(\eta^{\mu\rho}\partial^{\nu}\partial^{\sigma}
	+ \eta^{\mu\sigma}\partial^{\nu}\partial^{\rho}
	+ \eta^{\nu\rho}\partial^{\mu}\partial^{\sigma}
	+ \eta^{\nu\sigma}\partial^{\mu}\partial^{\rho}) \nonumber\\
& 	\hspace{1em}
	+ \frac{1}{2}(\eta^{\mu\nu}\partial^{\rho}\partial^{\sigma}
	+ \eta^{\rho\sigma}\partial^{\mu}\partial^{\nu}) \nonumber\\
& 	\hspace{1em}
	+ \frac{1}{4}
	(\eta^{\mu\rho}\eta^{\nu\sigma}
	+ \eta^{\mu\sigma}\eta^{\nu\rho}
	- 2\eta^{\mu\nu}\eta^{\rho\sigma})
	(\Box + (\partial_{z} + 3A')\partial_{z}), \nonumber\\
K^{\mu5; \rho5}
&= 	- \frac{1}{4}(\partial^{\mu}\partial^{\rho} - \eta^{\mu\rho}\Box), \nonumber\\
K^{\phi; \phi}
&= 	\frac{3}{4}\Box
	- \frac{3}{2}(\partial_{z} + A')(\partial_{z} + 2A'), \nonumber\\
K^{\mu\nu; \rho5}
&= 	- \frac{1}{4}
	(\eta^{\mu\rho}\partial^{\nu}
	+ \eta^{\nu\rho}\partial^{\mu}
	- 2\eta^{\mu\nu}\partial^{\rho})
	(\partial_{z} + 3A'), \nonumber\\
K^{\mu5; \rho\sigma}
&= 	- \frac{1}{4}
	(\eta^{\mu\rho}\partial^{\sigma}
	+ \eta^{\mu\sigma}\partial^{\rho}
	- 2\eta^{\rho\sigma}\partial^{\mu})
	\partial_{z}, \nonumber\\
K^{\mu\nu; \phi}
&= 	\frac{3}{4}\eta^{\mu\nu}(\partial_{z} + 3A')(\partial_{z} + 2A'), \nonumber\\
K^{\phi; \rho\sigma}
&= 	\frac{3}{4}\eta^{\rho\sigma}(\partial_{z} + A')\partial_{z}, \nonumber\\
K^{\mu5; \phi}
&= 	- \frac{3}{4}\partial^{\mu}(\partial_{z} + 2A'), \nonumber\\
K^{\phi; \rho5}
&= 	- \frac{3}{4}\partial^{\rho}(\partial_{z} + A'). \label{eq:gravity11}
\end{align}
\end{widetext}
This expression is consistent with the result given in \cite{GPP:2005} up to boundary terms, which are irrelevant in our discussions, and also with the equations of motion for the metric fluctuations in \cite{CGGP:2004}.

As we will see later, the general coordinate transformations are compatible with the mode expansions
\begin{align}
h_{\mu\nu}(x,z)
&= 	\sum_{n}h_{\mu\nu}^{(n)}(x)f^{(n)}(z), \nonumber\\
h_{\mu5}(x,z)
&= 	\sum_{n}h_{\mu5}^{(n)}(x)g^{(n)}(z), \nonumber\\
\phi(x,z)
&= 	\sum_{n}\phi^{(n)}(x)k^{(n)}(z) \label{eq:gravity12}
\end{align}
with a definite boundary condition for each fluctuation field.
The mode functions $f^{(n)}$, $g^{(n)}$, $k^{(n)}$ should be chosen to be the mass eigenstates or to diagonalize the quadratic action \eqref{eq:gravity10}.
It turns out that they have to satisfy the following Schr\"odinger-like equations:
\begin{align}
& 	- (\partial_{z}^{2} + 3A'\partial_{z})f^{(n)}(z) = m_{n}^{2}f^{(n)}(z), \label{eq:gravity13}\\
& 	- (\partial_{z}^{2} + 3A'\partial_{z} + 3A'')g^{(n)}(z) = m_{n}^{2}g^{(n)}(z), \label{eq:gravity14}\\
& 	- (\partial_{z}^{2} + 3A'\partial_{z} + 4A'')k^{(n)}(z) = m_{n}^{2}k^{(n)}(z). \label{eq:gravity15}
\end{align}
Surprisingly, we will find later that the mass eigenvalue $m_{n}$ is triply degenerate for $f^{(n)}$, $g^{(n)}$ and $k^{(n)}$ except for zero modes.
We should make a comment on delta-function potentials 
that would appear at the boundaries in the above equations.
Since we take an interval picture with two boundaries at
$z=z_{1}, z_{2}$,
the contribution of the delta-function potentials
should be absorbed into the boundary conditions for
$f^{(n)}(z), g^{(n)}(z)$ and $k^{(n)}(z)$, which will be
derived consistently from a supersymmetric point of view
in the subsection III.D.
\subsection{Supersymmetry} \label{sec:susy}
In this subsection, we show that a quantum-mechanical supersymmetric structure is hidden in the 4d spectrum.
To this end, let us first consider the eigenfunctions $f^{(n)}$ and $g^{(n)}$.
The supersymmetric structure will become apparent if we express the equations \eqref{eq:gravity13} and \eqref{eq:gravity14} into the form
\begin{align}
D^{\dagger}Df^{(n)}(z) &= m_{n}^{2}f^{(n)}(z), \nonumber\\
DD^{\dagger}g^{(n)}(z) &= m_{n}^{2}g^{(n)}(z) \label{eq:gravity16}
\end{align}
with
\begin{align}
D = \partial_{z}, \quad
D^{\dagger} = - (\partial_{z} + 3A'). \label{eq:gravity17}
\end{align}
If $D^{\dagger}$ is the hermitian conjugate to $D$, two functions $f^{(n)}$ and $g^{(n)}$ form an $N = 2$ supersymmetry multiplet.
This is indeed true with respect to the inner product
\begin{align}
\langle \psi | \varphi \rangle
= 	\int_{z_{1}}^{z_{2}}\!\!\!{\rm d}z~{\rm e}^{3A}
	\psi(z)^{*}\varphi(z) \label{eq:gravity18}
\end{align}
with the boundary conditions
\begin{align}
\partial_{z}f^{(n)}(z) = g^{(n)}(z) = 0 \quad
\text{at} \quad
z = z_{1}, z_{2}. \label{eq:gravity19}
\end{align}
The factor ${\rm e}^{3A}$ in Eq.\eqref{eq:gravity18} is required because of the presence of it in the action \eqref{eq:gravity10}, whose origin comes from the nontrivial metric \eqref{eq:gravity02}.
The boundary conditions \eqref{eq:gravity19} turn out to be compatible with supersymmetry and will be derived in the next subsection.

To rewrite the system into the $N = 2$ supersymmetric form, we introduce two-component vectors
\begin{align}
\Psi (z) = 	\begin{pmatrix}
		f(z) \\
		g(z)
		\end{pmatrix}, \label{eq:gravity20}
\end{align}
where they are assumed to obey the boundary conditions \eqref{eq:gravity19}.
The inner product of $\Psi_{1}(z)$ and $\Psi_{2}(z)$ is defined by
\begin{align}
\langle \Psi_{1} | \Psi_{2} \rangle
= 	\int_{z_{1}}^{z_{2}}\!\!\!{\rm d}z~{\rm e}^{3A}
	\bigl\{f_{1}^{*}(z)f_{2}(z) + g_{1}^{*}(z)g_{2}(z)\bigr\}. \label{eq:gravity21}
\end{align}
Then, the Hamiltonian and the supercharges are given by
\begin{align}
H = 	\begin{pmatrix}
	D^{\dagger}D 	& 0 \\
	0 			& DD^{\dagger}
	\end{pmatrix} \label{eq:gravity22}
\end{align}
and
\begin{align}
Q
= 	\begin{pmatrix}
	0 	& 0 \\
	D 	& 0
	\end{pmatrix}, \quad
Q^{\dagger}
= 	\begin{pmatrix}
	0 	& D^{\dagger} \\
	0 	& 0
	\end{pmatrix}. \label{eq:gravity23}
\end{align}
We further introduce the operator $(-1)^{F}$ with $F$ being the ``fermion'' number operator as
\begin{align}
(-1)^{F}
= 	\begin{pmatrix}
	1 	& 0 \\
	0 	& -1
	\end{pmatrix}. \label{eq:gravity24}
\end{align}
It is easy to show that the operators $H$, $Q$, $Q^{\dagger}$ and $(-1)^{F}$ form the $N = 2$ supersymmetry algebra
\begin{align}
& H = \{Q, Q^{\dagger}\}, \nonumber\\
& \{Q, Q\} = \{Q^{\dagger}, Q^{\dagger}\} = 0, \nonumber\\
& [Q, H] = [Q^{\dagger}, H] = 0, \nonumber\\
& [(-1)^{F}, H] = 0, \nonumber\\
& \{Q, (-1)^{F}\} = \{Q^{\dagger}, (-1)^{F}\} = 0. \label{eq;gravity25}
\end{align}
With respect to the inner product \eqref{eq:gravity21} and the boundary conditions \eqref{eq:gravity19}, $H$ and $(-1)^{F}$ are hermitian and $Q^{\dagger}$ is the hermitian conjugate to $Q$, and vice versa.
Since $(-1)^{F}$ commutes with $H$, we can have simultaneous eigenfunctions $\Psi_{\pm}^{(n)}$ of $H$ and $(-1)^{F}$ as
\begin{align}
H\Psi_{\pm}^{(n)} &= m_{n}^{2}\Psi_{\pm}^{(n)}, \nonumber\\
(-1)^{F}\Psi_{\pm}^{(n)} &= \pm\Psi_{\pm}^{(n)}. \label{eq:gravity26}
\end{align}
Because of supersymmetry, $\Psi_{+}^{(n)}$ is related to $\Psi_{-}^{(n)}$, with appropriate normalization, as
\begin{align}
Q\Psi_{+}^{(n)} &= m_{n}\Psi_{-}^{(n)}, \nonumber\\
Q^{\dagger}\Psi_{-}^{(n)} &= m_{n}\Psi_{+}^{(n)}. \label{eq:gravity27}
\end{align}
In terms of the component fields $f^{(n)}$ and $g^{(n)}$, we can write
\begin{align}
\Psi_{+}^{(n)}
= 	\begin{pmatrix}
	f^{(n)} \\
	0
	\end{pmatrix}, \quad
\Psi_{-}^{(n)}
= 	\begin{pmatrix}
	0 \\
	g^{(n)}
	\end{pmatrix} \label{eq:gravity28}
\end{align}
and
\begin{align}
Df^{(n)} &= m_{n}g^{(n)}, \nonumber\\
D^{\dagger}g^{(n)} &= m_{n}f^{(n)}. \label{eq:gravity29}
\end{align}
It follows that $f^{(n)}$ and $g^{(n)}$ (or $\Psi_{\pm}^{(n)}$) form a supersymmetric multiplet except for the zero mode with $m_{0} = 0$.

Let us next proceed to the analysis of a pair of the eigenfunctions $g^{(n)}$ and $k^{(n)}$.
One might expect that $g^{(n)}$ and $k^{(n)}$ could not form a supersymmetry multiplet because, if so, the mass eigenvalue $m_{n}$ is triply degenerate between $f^{(n)}$, $g^{(n)}$ and $k^{(n)}$ but supersymmetry allows only even numbers of degeneracy between ``bosonic'' and ``fermionic'' states.
Surprisingly, it turns out that $g^{(n)}$ and $k^{(n)}$ actually form a supersymmetry multiplet and that the spectrum can be described by another $N = 2$ supersymmetric quantum-mechanical one.
A key observation is that the second differential operator $- (\partial_{z}^{2} + 3A'\partial_{z} + 3A'')$ in Eq.\eqref{eq:gravity14} can be expressed in two supersymmetric ways:
\begin{align}
- (\partial_{z}^{2} + 3A'\partial_{z} + 3A'')
&= 	DD^{\dagger} \nonumber\\
&= 	{\bar D}^{\dagger}{\bar D}, \label{eq:gravity30}
\end{align}
where $D$ and $D^{\dagger}$ are defined in Eq.\eqref{eq:gravity17}, while ${\bar D}$ and ${\bar D}^{\dagger}$ are
\begin{align}
{\bar D} = \partial_{z} + A', \quad
{\bar D}^{\dagger} = - (\partial_{z} + 2A'). \label{eq:gravity31}
\end{align}
To verify the relations \eqref{eq:gravity30}, we will use the identity $(A')^{2} = A''$.
A crucial point is that ${\bar D}^{\dagger}$ is the hermitian conjugate to ${\bar D}$ with respect to the inner product \eqref{eq:gravity18} and the boundary conditions \eqref{eq:gravity19} for $g^{(n)}(z)$.
The relation \eqref{eq:gravity30}, however, seems strange because a zero mode $g^{(0)}$ with $m_{0} = 0$ has to satisfy both of the equations
\begin{align}
D^{\dagger}g^{(0)} = 0 \quad
\text{and} \quad
{\bar D}g^{(0)} = 0. \nonumber
\end{align}
This is impossible because $D^{\dagger}$ and ${\bar D}^{\dagger}$ are the first differential operators so that any (nontrivial) solution to $D^{\dagger}g^{(0)} = 0$ cannot satisfy the other equation ${\bar D}g^{(0)} = 0$, and vice versa.
A loophole in the above argument is that the eigenfunctions $g^{(n)}$ have no zero mode with $m_{0} = 0$.
The boundary conditions \eqref{eq:gravity19} for $g^{(n)}(z)$ actually forbid any nontrivial solution to $D^{\dagger}g^{(0)} = 0$ and ${\bar D}g^{(0)} = 0$.

The supersymmetric structure for $g^{(n)}$ and $k^{(n)}$ will become apparent if we express the equations \eqref{eq:gravity14} and \eqref{eq:gravity15} into the form
\begin{align}
{\bar D}^{\dagger}{\bar D}g^{(n)}(z) &= m_{n}^{2}g^{(n)}(z), \nonumber\\
{\bar D}{\bar D}^{\dagger}k^{(n)}(z) &= m_{n}^{2}k^{(n)}(z). \label{eq:gravity32}
\end{align}
To rewrite the system into the $N = 2$ supersymmetric form, we introduce two-component vectors
\begin{align}
\Phi(z) = 	\begin{pmatrix}
		g(z) \\
		k(z)
		\end{pmatrix} \label{eq:gravity33}
\end{align}
with the inner product
\begin{align}
\langle \Phi_{1} | \Phi_{2} \rangle
= 	\int_{z_{1}}^{z_{2}}\!\!\!{\rm d}z~{\rm e}^{3A}
	\bigl\{g_{1}^{*}(z)g_{2}(z) + k_{1}^{*}(z)k_{2}(z)\bigr\} \label{eq:gravity34}
\end{align}
with the boundary conditions
\begin{align}
g(z) = (\partial_{z} + 2A')k(z) = 0 \quad
\text{at} \quad
z = z_{1}, z_{2}. \label{eq:gravity35}
\end{align}
Then, the Hamiltonian and the supercharges are defined by
\begin{align}
{\bar H}
= 	\begin{pmatrix}
	{\bar D}^{\dagger}{\bar D} 	& 0 \\
	0 						& {\bar D}{\bar D}^{\dagger}
	\end{pmatrix}, \label{eq:gravity36}
\end{align}
and
\begin{align}
{\bar Q}
= 	\begin{pmatrix}
	0 		& 0 \\
	{\bar D} 	& 0
	\end{pmatrix}, \quad
{\bar Q}^{\dagger}
= 	\begin{pmatrix}
	0 	& {\bar D}^{\dagger} \\
	0 	& 0
	\end{pmatrix}. \label{eq:gravity37}
\end{align}
We further introduce the operator $(-1)^{\bar F}$ with ${\bar F}$ being the ``fermion'' number operator
\begin{align}
(-1)^{\bar F}
= 	\begin{pmatrix}
	1 	& 0 \\
	0 	& -1
	\end{pmatrix}. \label{eq:gravity38}
\end{align}
As before, the operators ${\bar H}$, ${\bar Q}$, ${\bar Q}^{\dagger}$, $(-1)^{\bar F}$ satisfy the $N = 2$ supersymmetry algebra.
The eigenfunctions $g^{(n)}$ and $k^{(n)}$ form a supersymmetry multiplet and are related each other, with appropriate normalization, as
\begin{align}
{\bar D}g^{(n)} &= m_{n}k^{(n)}, \nonumber\\
{\bar D}^{\dagger}k^{(n)} &= m_{n}g^{(n)}. \label{eq:gravity39}
\end{align}

We have shown that {\em two} $N = 2$ supersymmetric systems are hidden in the 4d spectrum of the 5d gravity theory.
It suggests that the system could realize some extension of the $N = 2$ supersymmetry.
One might expect that the two $N = 2$ supersymmetric systems could be embedded in an $N = 4$ supersymmetric one.
This is not, however, the case because three-fold degeneracy does not match the standard supersymmetry.
It may be necessary to search for some nonstandard extension of the $N = 2$ supersymmetry.
We would like to discuss this subject before closing this subsection.

To this end, let us consider a pair of the eigenfunctions $f^{(n)}$ and $k^{(n)}$ which obey the equations
\begin{align}
H_{f}f^{(n)} &= m_{n}^{2}f^{(n)}, \nonumber\\
H_{k}k^{(n)} &= m_{n}^{2}k^{(n)}, \label{eq:gravity40}
\end{align}
where
\begin{align}
H_{f}
&= 	D^{\dagger}D = - (\partial_{z} + 3A')\partial_{z}, \nonumber\\
H_{k}
&= 	{\bar D}{\bar D}^{\dagger} = - (\partial_{z} + A')(\partial_{z} + 2A'). \label{eq:gravity41}
\end{align}
It turns out that the Hamiltonians $H_{f}$ and $H_{k}$ are related each other through the so-called intertwining relation
\begin{align}
{\cal A}H_{f} = H_{k}{\cal A}, \label{eq:gravity42}
\end{align}
where the intertwiner ${\cal A}$ is given by
\begin{align}
{\cal A} = {\bar D}D = (\partial_{z} + A')\partial_{z}. \label{eq:gravity43}
\end{align}
If we introduce the following operators
\begin{align}
{\cal H}
&= 	\begin{pmatrix}
	H_{f} 	& 0 \\
	0 		& H_{k}
	\end{pmatrix}, \nonumber\\
{\cal Q}
&= 	\begin{pmatrix}
	0 		& 0 \\
	{\cal A} 	& 0
	\end{pmatrix}, \quad
{\cal Q}^{\dagger}
= 	\begin{pmatrix}
	0 	& {\cal A}^{\dagger} \\
	0 	& 0
	\end{pmatrix}, \label{eq:gravity44}
\end{align}
we then find the interesting relations
\begin{align}
& \{{\cal Q}, {\cal Q}^{\dagger}\} = {\cal H}^{2}, \nonumber\\
& \{{\cal Q}, {\cal Q}\} = \{{\cal Q}^{\dagger}, {\cal Q}^{\dagger}\} = 0, \nonumber\\
& [{\cal Q}, {\cal H}] = [{\cal Q}^{\dagger}, {\cal H}] = 0. \label{eq:gravity45}
\end{align}
It should be emphasized that ${\cal Q}$ and ${\cal Q}^{\dagger}$ are the second order differential operators, as opposed to the ordinary supercharges, and that the left-hand-side of the first equation in Eq.\eqref{eq:gravity45} is given by the square of the Hamiltonian ${\cal H}$ but not the linear of it.
The system with the nonlinear algebra has been discussed in \cite{ACDI:1995,AIN:1995,Fernandez:1997,AIN:1999,AST:2001,CIN:2002,AS:2003} as an extension of the $N = 2$ supersymmetry.

Another type of extensions of the $N = 2$ supersymmetry has also been discussed in \cite{LNOSS3:2007}.
The system may be characterized by the following nonlinear relations:
\begin{align}
& 	{\cal H}^{{\cal N}-1} = {\cal Q}^{\cal N} = ({\cal Q}^{\dagger})^{\cal N}, \nonumber\\
& 	\Omega^{\cal N} = \bf{1}, \nonumber\\
& 	\Omega{\cal Q} = \omega{\cal Q}\Omega, \quad
	\Omega{\cal Q}^{\dagger} = \omega^{-1}{\cal Q}^{\dagger}\Omega, \quad
	\omega = {\rm e}^{2\pi i/{\cal N}}, \nonumber\\
& 	[{\cal Q} , {\cal H}] = [{\cal Q}^{\dagger}, {\cal H}] = 0, \nonumber\\
& 	[\Omega, {\cal H}] = 0. \label{eq:gravity46}
\end{align}
The system is shown to be ${\cal N}$-fold degenerate except for zero modes.
We note that the above nonlinear algebra reduces to the original $N = 2$ supersymmetry algebra when ${\cal N} = 2$.
Our gravitational system corresponds to ${\cal N} = 3$ and the operators in Eq.\eqref{eq:gravity46} can be realized as
\begin{align}
{\cal H}
&= 	\begin{pmatrix}
	D^{\dagger}D 	& 0 				& 0 \\
	0 			& DD^{\dagger} 	& 0 \\
	0 			& 0				& {\bar D}{\bar D}^{\dagger}
	\end{pmatrix}
= 	\begin{pmatrix}
	D^{\dagger}D 	& 0 						& 0 \\
	0 			& {\bar D}^{\dagger}{\bar D} 	& 0 \\
	0 			& 0						& {\bar D}{\bar D}^{\dagger}
	\end{pmatrix}, \nonumber\\
{\cal Q}
&= 	\begin{pmatrix}
	0 	& 0 			& D^{\dagger}{\bar D}^{\dagger} \\
	D 	& 0 			& 0 \\
	0 	& {\bar D} 	& 0
	\end{pmatrix}, \quad
{\cal Q}^{\dagger}
= 	\begin{pmatrix}
	0 			& D^{\dagger} 	& 0 \\
	0 			& 0 			& {\bar D}^{\dagger} \\
	{\bar D}D 	& 0 			& 0
	\end{pmatrix}, \nonumber\\
\Omega
&= 	\begin{pmatrix}
	1 	& 0 			& 0 \\
	0 	& \omega 	& 0 \\
	0 	& 0 			& \omega^{2}
	\end{pmatrix}, \quad
\omega = {\rm e}^{2\pi i/ 3}. \label{eq:gravity47}
\end{align}
The above operators act on three-component vectors
\begin{align}
\Psi(x)
= 	\begin{pmatrix}
	f(z) \\
	g(z) \\
	k(z)
	\end{pmatrix}. \label{eq:gravity48}
\end{align}
Since ${\cal H}$ and $\Omega$ commute each other, we can have simultaneous eigenfunctions of them such as
\begin{align}
{\cal H}\Psi_{\omega^{l}}^{(n)} &= m_{n}^{2}\Psi_{\omega^{l}}^{(n)}, \nonumber\\
\Omega\Psi_{\omega^{l}}^{(n)} &= \omega^{l}\Psi_{\omega^{l}}^{(n)}, \quad
l = 0, 1, 2. \label{eq:gravity49}
\end{align}
The relations $[{\cal Q}, {\cal H}] = [{\cal Q}^{\dagger}, {\cal H}] = 0$ and $\Omega{\cal Q} = \omega{\cal Q}\Omega$, $\Omega{\cal Q}^{\dagger} = \omega^{-1}{\cal Q}^{\dagger}\Omega$ imply that
\begin{align}
{\cal Q}\Psi_{\omega^{l}}^{(n)} &\propto \Psi_{\omega^{l+1}}^{(n)}, \nonumber\\
{\cal Q}^{\dagger}\Psi_{\omega^{l}}^{(n)} &\propto \Psi_{\omega^{l-1}}^{(n)}. \label{eq:gravity50}
\end{align}
It follows that the spectrum is triply degenerate (except for zero modes).
In terms of the eigenfunctions $f^{(n)}$, $g^{(n)}$ and $k^{(n)}$, $\Psi_{\omega^{l}}^{(n)}$ are explicitly given by
\begin{align}
\Psi_{1}^{(n)}
= 	\begin{pmatrix}
	f^{(n)} \\
	0 \\
	0
	\end{pmatrix}, \quad
\Psi_{\omega}^{(n)}
= 	\begin{pmatrix}
	0 \\
	g^{(n)} \\
	0
	\end{pmatrix}, \quad
\Psi_{\omega^{2}}^{(n)}
= 	\begin{pmatrix}
	0 \\
	0 \\
	k^{(n)}
	\end{pmatrix}. \label{eq:gravity51}
\end{align}
\subsection{Boundary Conditions} \label{sec:BC}
We have assumed in the previous subsection that the boundary conditions of $f^{(n)}$, $g^{(n)}$, $k^{(n)}$ are given by
\begin{align}
& 	\partial_{z}f^{(n)}(z) = 0, \nonumber\\
& 	g^{(n)}(z) = 0, \nonumber\\
& 	(\partial_{z} + 2A')k^{(n)}(z) = 0, \quad
	\text{at} \quad
	z = z_{1}, z_{2}, \label{eq:gravity52}
\end{align}
which imply that the original fluctuation fields have to obey
\begin{align}
& 	\partial_{z}h_{\mu\nu}(x,z) = 0, \nonumber\\
& 	h_{\mu5}(x,z) = 0, \nonumber\\
& 	(\partial_{z} + 2A')\phi(x,z) = 0, \quad
	\text{at} \quad
	z = z_{1}, z_{2}. \label{eq:gravity53}
\end{align}
In five-dimensional braneworld gravity, two approaches have been proposed to obtain the boundary conditions.
The first is to impose the $\mathbb{Z}_{2}$ orbifold symmetry to the equations of motion, and simply integrate them around the neighborhood of orbifold fixed points to obtain the junction conditions \cite{CGR:2000,GRS:2000,PRZ:2000,DL:2003,GPP:2005}.
This is the most convenient way to obtain boundary conditions consistent with Israel junction condition \cite{Israel:1966}.
It, however, seems not to be applicable to the interval picture (not the $\mathbb{Z}_{2}$ orbifold picture), 
where the extra dimension is limited to the space between two branes
and $\mathbb{Z}_{2}$ symmetry is meaningless.
The second is to introduce the Gibbons-Hawking extrinsic curvature terms \cite{GH:1977} on branes and then obtain the boundary conditions by the variational principle \cite{LM:2001,CLP:2005,BCLPS:2006}.
Since the boundary conditions are crucially important to determine the spectrum, especially zero mode, it will be worthwhile deriving them from various different points of view.
In this subsection, we propose the third approach to derive them from a supersymmetric point of view, which is quite different from other geometrical approaches.

Let us start with the differential operator defined by
\begin{align}
H_{\gamma}
= - \bigl(\partial_{z} + (3 - \gamma)A'\bigr)\bigl(\partial_{z} + \gamma A'\bigr), \label{eq:gravity54}
\end{align}
where $H_{\gamma}$ corresponds to the Hamiltonians for $f^{(n)}$, $g^{(n)}$ and $k^{(n)}$ with $\gamma = 0$, $3$(or $1$) and $2$, respectively.
We then require $H_{\gamma}$ to be hermitian, i.e.
\begin{align}
\langle \psi | H_{\gamma}\varphi \rangle
= \langle H_{\gamma}\psi | \varphi \rangle \label{eq:gravity55}
\end{align}
for any functions $\psi$ and $\varphi$ obeying appropriate boundary conditions.
The hermiticity of $H_{\gamma}$ is found to be assured if $\psi$ and $\varphi$ satisfy
\begin{align}
\bigl(\psi(z)\bigr)^{*}\partial_{z}\varphi(z) - \bigl(\partial_{z}\psi(z)\bigr)^{*}\varphi(z) = 0 \quad
\text{at} \quad
z = z_{1}, z_{2}. \label{eq:gravity56}
\end{align}
The conditions can be realized only if $\psi(z)$ (and also $\varphi(z)$) obeys the following boundary conditions:
\footnote
{%
Here, we have assumed that $\psi(z_{1})$ and $\varphi(z_{1})$ are independent of $\psi(z_{2})$ and $\varphi(z_{2})$ because the extra dimension is an interval with two boundaries.
If we allow them to relate to each other, we would have a wider class of possible boundary conditions \cite{NST:2003,NST:2004,NST:2005}.
}
\begin{align}
\kappa\cos\theta_{i}\psi(z_{i}) = \sin\theta_{i}\partial_{z}\psi(z_{i}), \quad
i = 1, 2, \label{eq:gravity57}
\end{align}
where $\theta_{i}$ ($i = 1$, $2$) are arbitrary real constants
and $\kappa$ is a nonzero real constant of mass dimension one,
which is introduced to adjust the mass dimension of Eq.(\ref{eq:gravity57}).
The above result implies that the functions $f^{(n)}$, $g^{(n)}$ and $k^{(n)}$ have to obey
\begin{align}
\kappa\cos\theta_{i}^{f}f^{(n)}(z_{i})
&= 	\sin\theta_{i}^{f}\partial_{z}f^{(n)}(z_{i}), \label{eq:gravity58}\\
\kappa\cos\theta_{i}^{g}g^{(n)}(z_{i})
&= 	\sin\theta_{i}^{g}\partial_{z}g^{(n)}(z_{i}), \label{eq:gravity59}\\
\kappa\cos\theta_{i}^{k}k^{(n)}(z_{i})
&= 	\sin\theta_{i}^{k}\partial_{z}k^{(n)}(z_{i}), \quad
	i = 1, 2, \label{eq:gravity60}
\end{align}
for some real constants $\theta_{i}^{f}$, $\theta_{i}^{g}$, $\theta_{i}^{k}$ ($i = 1$, $2$).

As was shown in the previous subsection, the eigenvalue $m_{n}$ for $f^{(n)}$, $g^{(n)}$ and $k^{(n)}$ is three-fold degenerate.
The degeneracy would not, however, hold for general values of $\theta_{i}^{f}$, $\theta_{i}^{g}$ and $\theta_{i}^{k}$ because the supersymmetric relations between $f^{(n)}$, $g^{(n)}$ and $k^{(n)}$
\begin{align}
\partial_{z}f^{(n)}(z)
&= 	m_{n}g^{(n)}(z), \label{eq:gravity61}\\
- (\partial_{z} + 3A')g^{(n)}(z)
&= 	m_{n}f^{(n)}(z), \label{eq:gravity62}\\
(\partial_{z} + A')g^{(n)}(z)
&= 	m_{n}k^{(n)}(z), \label{eq:gravity63}\\
- (\partial_{z} + 2A')k^{(n)}(z)
&= 	m_{n}g^{(n)}(z) \label{eq:gravity64}
\end{align}
are generally inconsistent with the conditions \eqref{eq:gravity58} -- \eqref{eq:gravity60}.
We should emphasize that the relations \eqref{eq:gravity61} -- \eqref{eq:gravity64} guarantee the degeneracy of the eigenvalues for $f^{(n)}$, $g^{(n)}$ and $k^{(n)}$.
Using Eqs.\eqref{eq:gravity58}, \eqref{eq:gravity61}, \eqref{eq:gravity62}, we find
\begin{align}
- (3A'\kappa\cos\theta_{i}^{f} + m_{n}^{2}\sin\theta_{i}^{f})g^{(n)}(z_{i})
&= 	\kappa\cos\theta_{i}^{f}\partial_{z}g^{(n)}(z_{i}), \nonumber\\
&	\hspace{6em}
	i = 1, 2. \label{eq:gravity65}
\end{align}
The boundary conditions for $g^{(n)}(z)$ have to be independent of $n$, otherwise the superposition of $g^{(n)}(z)$ would be meaningless.
It follows that
\begin{align}
\sin\theta_{i}^{f} = 0 \quad
\text{or} \quad
\cos\theta_{i}^{f} = 0, \quad
i = 1, 2. \label{eq:gravity66}
\end{align}
The conditions lead to
\begin{align}
\partial_{z}f^{(n)}(z_{i}) = 0 \quad
\text{and} \quad
g^{(n)}(z_{i}) = 0, \label{eq:gravity67}
\end{align}
or
\begin{align}
f^{(n)}(z_{i}) = 0 \quad
\text{and} \quad
(\partial_{z} + 3A')g^{(n)}(z_{i}) = 0, \quad
i = 1, 2. \label{eq:gravity68}
\end{align}
It may be instructive to note that the above conditions assure that $\Psi(z) = \begin{pmatrix} f^{(n)}(z) \\ g^{(n)}(z) \end{pmatrix}$ satisfies the same boundary conditions as $Q\Psi(z)$ and $Q^{\dagger}\Psi(z)$.
In other words, the supercharges $Q$ and $Q^{\dagger}$ act well-definedly on the functional space of $\Psi(z)$, as they should do.

We can repeat the same argument for $g^{(n)}(z)$ and $k^{(n)}(z)$.
For the eigenvalue $m_{n}$ of $g^{(n)}$ to be identical to that of $k^{(n)}$, they have to be related each other through the equations \eqref{eq:gravity63} and \eqref{eq:gravity64}.
We have already shown that the hermiticity of the Hamiltonians for $g^{(n)}$ and $k^{(n)}$ requires Eqs.\eqref{eq:gravity59} and \eqref{eq:gravity60}.
Those equations are compatible with the supersymmetric relations \eqref{eq:gravity63} and \eqref{eq:gravity64} only when
\begin{align}
g^{(n)}(z_{i}) = 0 \quad
\text{and} \quad
(\partial_{z} + 2A')k^{(n)}(z_{i}) = 0, \label{eq:gravity69}
\end{align}
or
\begin{align}
(\partial_{z} + A')g^{(n)}(z_{i}) = 0 \quad
\text{and} \quad
k^{(n)}(z_{i}) = 0, \quad
i = 1, 2. \label{eq:gravity70}
\end{align}
These conditions again insure that the supercharges ${\bar Q}$ and ${\bar Q}^{\dagger}$ act well-definedly on the functions $\Phi(z) = \begin{pmatrix} g^{(n)}(z) \\ k^{(n)}(z) \end{pmatrix}$, as they should do.

We have thus shown that candidates of possible boundary conditions are given by Eq.\eqref{eq:gravity67} or \eqref{eq:gravity68}, and Eq.\eqref{eq:gravity69} or \eqref{eq:gravity70}.
Each of the three combinations, Eqs.\eqref{eq:gravity67} and \eqref{eq:gravity70}, Eqs.\eqref{eq:gravity68} and \eqref{eq:gravity69}, Eqs.\eqref{eq:gravity68} and \eqref{eq:gravity70}, is, however, incompatible each other.
Hence, we finally arrive at the allowed boundary conditions \eqref{eq:gravity52} compatible with the supersymmetry, as announced before.
It is interesting to note that for the 5d gauge theory discussed in the previous section there are {\em four} types of possible boundary conditions compatible with the $N = 2$ supersymmetry.
On the other hand, for the 5d gravity theory, the boundary conditions are {\em uniquely} determined due to the existence of the two systems with the $N = 2$ supersymmetry.
\subsection{Spectrum} \label{sec:spectrum}
In the previous subsections, we have discussed the mode expansions and the boundary conditions for the metric fluctuation fields.
Although a number of studies have already been made on the 4d spectrum of the model, most of the works have concentrated on the physical spectrum by taking gauge fixing to remove unphysical degrees of freedom.
In this subsection, we present the quadratic action for the full KK modes {without gauge fixing}, from which we can clearly know how to take the unitary gauge to express the action in terms of the physical degrees of freedom.

Let us first consider the metric fluctuation field $h_{\mu\nu}(x,z)$.
The mode expansion of $h_{\mu\nu}(x,z)$ is given by
\begin{align}
h_{\mu\nu}(x,z)
= \sum_{n=0}^{\infty}h_{\mu\nu}^{(n)}(x)f^{(n)}(z), \label{eq:gravity71}
\end{align}
where the mass eigenfunctions $f^{(n)}$ are defined by
\begin{align}
- (\partial_{z} + 3A')\partial_{z}f^{(n)}(z) = m_{n}^{2}f^{(n)}(z) \label{eq:gravity72}
\end{align}
with the boundary conditions
\begin{align}
\partial_{z}f^{(n)}(z) = 0 \quad
\text{at} \quad
z = z_{1}, z_{2}. \label{eq:gravity73}
\end{align}
It follows that the boundary conditions \eqref{eq:gravity73} allow $f^{(n)}$ to have a zero mode, i.e.
\begin{align}
\partial_{z}f^{(0)}(z) = 0 \quad
\rightarrow \quad
f^{(0)} = {\rm const}, \label{eq:gravity74}
\end{align}
with $m_{0} = 0$.
The existence of the zero mode implies a massless graviton $h_{\mu\nu}^{(0)}(x)$.
The general solutions 
for $n \ne 0$
to the equation \eqref{eq:gravity72} with the boundary conditions \eqref{eq:gravity73} are found to be of the form
\begin{align}
f^{(n)}(z)
= 	C^{(n)}z^{2}
	\bigl({\rm Y}_{1}(m_{n}z_{2}){\rm J}_{2}(m_{n}z)
	- {\rm J}_{1}(m_{n}z_{2}){\rm Y}_{2}(m_{n}z)\bigr), \label{eq:gravity75}
\end{align}
where ${\rm J}_{\nu}$ is the Bessel function of the first kind of order $\nu$ and ${\rm Y}_{\nu}$ is the Bessel function of the second kind (or the Neumann function) of order $\nu$.
The $C^{(n)}$ is the (real) normalization constant which will be determined by
\begin{align}
M^{3}\int_{z_{1}}^{z_{2}}\!\!\!{\rm d}z~{\rm e}^{3A}
f^{(m)}(z)f^{(n)}(z) = \delta_{mn}M_{\rm Pl}^{2}, \label{eq:gravity76}
\end{align}
where
\begin{align}
M_{\rm Pl}^{2}
= 	M^{3}\int_{z_{1}}^{z_{2}}\!\!\!{\rm d}z~{\rm e}^{3A}
	\bigl(f^{(0)}\bigr)^{2}. \nonumber
\end{align}
The mass eigenvalues $m_{n}$ are obtained from the solutions to the equation
\begin{align}
{\rm Y}_{1}(m_{n}z_{2}){\rm J}_{1}(m_{n}z_{1})
- {\rm J}_{1}(m_{n}z_{2}){\rm Y}_{1}(m_{n}z_{1}) = 0. \label{eq:gravity77}
\end{align}

Let us next consider the metric fluctuation field $h_{\mu5}(x,z)$.
The mode expansion of $h_{\mu5}(x,z)$ is given by
\begin{align}
h_{\mu5}(x,z) = \sum_{n=1}^{\infty}h_{\mu5}^{(n)}(x)g^{(n)}(z). \label{eq:gravity78}
\end{align}
Thanks to supersymmetry, the function $g^{(n)}(z)$ can be obtained from $f^{(n)}(z)$ through the relation
\begin{align}
g^{(n)}(z)
&= 	\frac{1}{m_{n}}\partial_{z}f^{(n)}(z) \nonumber\\
&= 	C^{(n)}z^{2}
	\bigl[{\rm Y}_{1}(m_{n}z_{2}){\rm J}_{1}(m_{n}z)
	- {\rm J}_{1}(m_{n}z_{2}){\rm Y}_{1}(m_{n}z)\bigr]. \label{eq:gravity79}
\end{align}
As was noticed before, there is no zero mode for $g^{(n)}$ because a would-be zero mode solution does not satisfy the boundary conditions and hence the mode has to be removed from the spectrum.

\begin{figure}[t]
	\begin{center}
	\includegraphics[scale=.7]{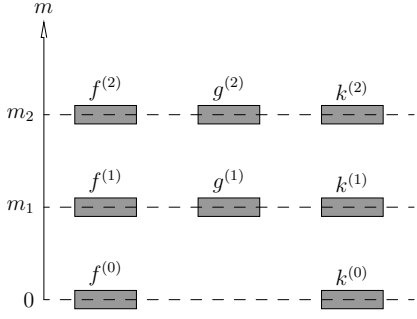}\\
	\caption
	{%
	The mass spectrum of $f^{(n)}$, $g^{(n)}$ and $k^{(n)}$.
	The spectrum is three-fold degenerate except for the zero modes.
	}
	\label{fig:spectrum2}
	\end{center}
\end{figure}

Let us finally discuss the metric fluctuation field $\phi(x,z)$.
The mode expansion of $\phi(x,z)$ is given by
\begin{align}
\phi(x,z)
= \sum_{n=0}^{\infty}\phi^{(n)}(x)k^{(n)}(z), \label{eq:gravity80}
\end{align}
where the mass eigenfunctions $k^{(n)}(z)$ are defined by
\begin{align}
- (\partial_{z} + A')(\partial_{z} + 2A')k^{(n)}(z) = m_{n}^{2}k^{(n)}(z) \label{eq:gravity81}
\end{align}
with the boundary conditions
\begin{align}
(\partial_{z} + 2A')k^{(n)}(z) = 0 \quad
\text{at} \quad
z = z_{1}, z_{2}. \label{eq:gravity82}
\end{align}
Again thanks to supersymmetry, the functions $k^{(n)}(z)$ with $n\neq0$ can be obtained from $f^{(n)}(z)$ through the relation
\begin{align}
k^{(n)}(z)
&= 	\frac{1}{m_{n}}(\partial_{z} + A')g^{(n)}(z) \nonumber\\
&= 	\frac{1}{m_{n}^{2}}(\partial_{z} + A')\partial_{z}f^{(n)}(z) \nonumber\\
&= 	C^{(n)}z^{2}
	\bigl[{\rm Y}_{1}(m_{n}z_{2}){\rm J}_{0}(m_{n}z)
	- {\rm J}_{1}(m_{n}z_{2}){\rm Y}_{0}(m_{n}z)\bigr]. \label{eq:gravity83}
\end{align}
The zero mode $k^{(0)}(z)$ is given by the solution of the first order differential equation
\begin{align}
(\partial_{z} + 2A')k^{(0)}(z) = 0 \quad
\rightarrow \quad
k^{(0)}(z) = \alpha^{(0)}z^{2}, \label{eq:gravity84}
\end{align}
which is consistent with the boundary conditions \eqref{eq:gravity82}.
The existence of the zero mode $k^{(0)}$ implies a massless scalar $\phi^{(0)}(x)$, which is called a radion.
A typical spectrum of the functions $f^{(n)}$, $g^{(n)}$ and $k^{(n)}$ is depicted in FIG. \ref{fig:spectrum2}.

\begin{widetext}
Inserting the mode expansions \eqref{eq:gravity71}, \eqref{eq:gravity78} and \eqref{eq:gravity80} into the quadratic action \eqref{eq:gravity10} and integrating out over the $z$ coordinate, we have
\begin{align}
S^{(2)}\bigl(h_{\mu\nu}^{(n)}, h_{\mu5}^{(n)}, \phi^{(n)}\bigr)
&= 	M_{\rm Pl}^{2}\int{\rm d}^{4}x
	\Biggl\{
	\sum_{n=0}^{\infty}\frac{1}{2}h_{\mu\nu}^{(n)}(x)
	\biggl[
	- \frac{1}{4}
	(\eta^{\mu\rho}\partial^{\nu}\partial^{\sigma}
	+ \eta^{\mu\sigma}\partial^{\nu}\partial^{\rho}
	+ \eta^{\nu\rho}\partial^{\mu}\partial^{\sigma}
	+ \eta^{\nu\sigma}\partial^{\mu}\partial^{\rho}) \nonumber\\
& 	\hspace{11em}
	+ \frac{1}{2}
	  (\eta^{\mu\nu}\partial^{\rho}\partial^{\sigma}
	+ \eta^{\rho\sigma}\partial^{\mu}\partial^{\nu}) \nonumber\\
&	\hspace{11em}
	+ \frac{1}{4}
	(\eta^{\mu\rho}\eta^{\nu\sigma}
	+ \eta^{\mu\sigma}\eta^{\nu\rho}
	- 2\eta^{\mu\nu}\eta^{\rho\sigma})
	(\Box - m_{n}^{2})\biggr]
	h_{\rho\sigma}^{(n)}(x) \nonumber\\
& 	\hspace{5em}
	+ \sum_{n=1}^{\infty}
	\frac{1}{2}h_{\mu5}^{(n)}(x)
	\bigl[
	\eta^{\mu\rho}\Box - \partial^{\mu}\partial^{\rho}\bigr]
	h_{\rho5}^{(n)}(x) \nonumber\\
&	\hspace{5em}
	+ \sum_{n=0}^{\infty}
	\frac{3}{8}\phi^{(n)}(x)
	\bigl[
	\Box + 2m_{n}^{2}\bigl]
	\phi^{(n)}(x) \nonumber\\
& 	\hspace{5em}
	+ \sum_{n=1}^{\infty}
	\frac{1}{4}
	\biggl(
	h_{\mu\nu}^{(n)}(x)
	\bigl[
	m_{n}(\eta^{\mu\rho}\partial^{\nu}
	+ \eta^{\nu\rho}\partial^{\mu}
	- 2\eta^{\mu\nu}\partial^{\rho})\bigr]
	h_{\rho5}^{(n)}(x) \nonumber\\
& 	\hspace{8em}
	+ h_{\mu5}^{(n)}(x)
	\bigl[
	- m_{n}(\eta^{\mu\rho}\partial^{\sigma}
	+ \eta^{\mu\sigma}\partial^{\rho}
	- 2\eta^{\rho\sigma}\partial^{\mu})\bigr]
	h_{\rho\sigma}^{(n)}(x)\biggr) \nonumber\\
& 	\hspace{5em}
	+ \sum_{n=0}^{\infty}
	\biggl(
	h_{\mu}^{(n)\mu}(x)
	\left[
	 \frac{3}{8}m_{n}^{2}\right]
	\phi^{(n)}(x)
	+ \phi^{(n)}(x)
	\left[
	 \frac{3}{8}m_{n}^{2}\right]
	h_{\rho}^{(n)\rho}(x)\biggr) \nonumber\\
& 	\hspace{5em}
	+ \sum_{n=1}^{\infty}
	\biggl(
	h_{\mu5}^{(n)}(x)
	\left[\frac{3}{4}m_{n}\partial^{\mu}\right]
	\phi^{(n)}(x)
	+ \phi^{(n)}(x)
	\left[ - \frac{3}{4}m_{n}\partial^{\rho}\right]
	h_{\rho5}^{(n)}\biggr)\Biggr\}. \label{eq:gravity85}
\end{align}
\end{widetext}
To clarify the physical degrees of freedom in the 4d spectrum, we would like to take the unitary gauge, in which all unphysical modes are gauged away from the action and only physical degrees of freedom survive.
To this end, let us recall linearized general coordinate transformations in terms of the KK modes:
\begin{align}
h_{\mu\nu}^{(n)}(x)
&\to 	{\hat h}_{\mu\nu}^{(n)}(x)
	= h_{\mu\nu}^{(n)}(x) - \partial_{\mu}\xi_{\nu}^{(n)}(x) - \partial_{\nu}\xi_{\mu}^{(n)}(x) \nonumber\\
&	\hspace{5em}
	+ m_{n}\xi_{5}^{(n)}(x)\eta_{\mu\nu}, \quad
	n = 0, 1, 2, \cdots, \nonumber\\
h_{\mu5}^{(n)}(x)
&\to 	{\hat h}_{\mu5}^{(n)}(x)
	= h_{\mu5}^{(n)}(x) - m_{n}\xi_{\mu}^{(n)}(x) - \partial_{\mu}\xi_{5}^{(n)}(x), \nonumber\\
&	\hspace{13em}
	n = 1, 2, 3, \cdots, \nonumber\\
\phi^{(n)}(x)
&\to 	{\hat \phi}^{(n)}(x)
	= \phi^{(n)}(x) - 2m_{n}\xi_{5}^{(n)}(x), \quad
	n = 0, 1, 2, \cdots. \label{eq:gravity86}
\end{align}
Here, we have expanded the gauge parameters $\xi_{\mu}(x,z)$ and $\xi_{5}(x,z)$ as
\begin{align}
\xi_{\mu}(x,z)
&= 	\sum_{n=0}^{\infty}\xi_{\mu}^{(n)}(x)f^{(n)}(z), \nonumber\\
\xi_{5}(x,z)
&= 	\sum_{n=1}^{\infty}\xi_{5}^{(n)}(x)g^{(n)}(z). \label{eq:gravity87}
\end{align}
It follows that we can take the unitary gauge
\begin{align}
{\hat h}_{\mu5}^{(n)}(x) &= 0, \quad n = 1, 2, 3, \cdots, \nonumber\\
{\hat \phi}^{(n)}(x) &= 0, \quad n = 1, 2, 3, \cdots, \label{eq:gravity88}
\end{align}
with the choice
\begin{align}
\xi_{\mu}^{(n)}(x)
&= 	\frac{1}{m_{n}}
	\left(h_{\mu5}^{(n)}(x) - \frac{1}{2m_{n}}\partial_{\mu}\phi^{(n)}(x)\right), \quad
	n = 1, 2, 3, \cdots, \nonumber\\
\xi_{5}^{(n)}(x)
&= 	\frac{1}{2m_{n}}\phi^{(n)}(x), \quad
	n = 1, 2, 3, \cdots. \label{eq:gravity89}
\end{align}
We should note that the zero mode $\phi^{(0)}$ cannot be gauged away and is physical.
Therefore, we conclude that the physical degrees of freedom are given by ${\hat h}_{\mu\nu}^{(n)}$ ($n = 0, 1, 2, \cdots$) and ${\hat \phi}^{(0)}$; ${\hat h}_{\mu\nu}^{(0)}$ is a massless graviton, ${\hat \phi}^{(0)}$ is a real massless scalar that is called a radion, and ${\hat h}_{\mu\nu}^{(n)}$ ($n = 1, 2, 3, \cdots$) are the massive gravitons that can become massive by eating the unphysical modes $h_{\mu5}^{(n)}$ and $\phi^{(n)}$ ($n = 1, 2, 3, \cdots$).

\begin{widetext}
An alternative way to obtain the physical degrees of freedom is to rewrite the quadratic action \eqref{eq:gravity85} into the following simple form
\begin{align}
S^{(2)}\bigl(h_{\mu\nu}^{(n)}, h_{\mu5}^{(n)}, \phi^{(n)}\bigr)
&= 	M_{\rm Pl}^{2}\int{\rm d}^{4}x
	\left\{
	\sum_{n=0}^{\infty}\frac{1}{2}{\tilde h}_{\mu\nu}^{(n)}(x)K^{\mu\nu; \rho\sigma}_{(n)}{\tilde h}_{\rho\sigma}^{(n)}(x)
	+ \frac{3}{8}\phi^{(0)}(x)\Box\phi^{(0)}(x)\right\} \nonumber\\
&= 	S^{(2)}\bigl({\tilde h}_{\mu\nu}^{(n)}, h_{\mu5}^{(n)} = 0, \phi^{(0)}, \phi^{(n\neq0)} = 0\bigr), \label{eq:gravity90}
\end{align}
where
\begin{align}
K^{\mu\nu; \rho\sigma}_{(n)}
&= 	- \frac{1}{4}
	(\eta^{\mu\rho}\partial^{\nu}\partial^{\sigma}
	+ \eta^{\mu\sigma}\partial^{\nu}\partial^{\rho}
	+ \eta^{\nu\rho}\partial^{\mu}\partial^{\sigma}
	+ \eta^{\nu\sigma}\partial^{\mu}\partial^{\rho}) \nonumber\\
& 	\hspace{1em}
	+ \frac{1}{2}(\eta^{\mu\nu}\partial^{\rho}\partial^{\sigma}
	+ \eta^{\rho\sigma}\partial^{\mu}\partial^{\nu} ) \nonumber\\
&	\hspace{1em}
	+ \frac{1}{4}
	(\eta^{\mu\rho}\eta^{\nu\sigma}
	+ \eta^{\mu\sigma}\eta^{\nu\rho}
	- 2\eta^{\mu\nu}\eta^{\rho\sigma})
	(\Box - m_{n}^{2}), \nonumber\\
{\tilde h}_{\mu\nu}^{(n)}
&= 	h_{\mu\nu}^{(n)}
	- \frac{1}{m_{n}}
	\left(\partial_{\mu}h_{\nu5}^{(n)} + \partial_{\nu}h_{\mu5}^{(n)}
	- \frac{1}{m_{n}}\partial_{\mu}\partial_{\nu}\phi^{(n)}\right)
	+ \frac{1}{2}\eta_{\mu\nu}\phi^{(n)}, \quad
	n = 1, 2, 3, \cdots, \nonumber\\
{\tilde h}_{\mu\nu}^{(0)}
&= 	h_{\mu\nu}^{(0)}. \label{eq:gravity91}
\end{align}
\end{widetext}
It is now clear that the action \eqref{eq:gravity90} (without gauge fixing) depends only on the fields ${\tilde h}_{\mu\nu}^{(n)}$ ($n = 0, 1, 2, \cdots$) and $\phi^{(0)}$.
Furthermore, ${\tilde h}_{\mu\nu}^{(n)}$ ($n = 1, 2, 3, \cdots$) and $\phi^{(0)}$ are invariant under the general coordinate transformations \eqref{eq:gravity86} and hence they are physical degrees of freedom.
The zero mode ${\tilde h}_{\mu\nu}^{(0)}$ is not invariant under the transformation \eqref{eq:gravity86} because the action remains invariant under the 4d general coordinate transformations.
\section{CONCLUSIONS AND DISCUSSIONS} \label{sec:concl}
In this paper, we have investigated the 5d gravity theory with the Randall-Sundrum background metric without taking any gauge fixing and shown that the 4d spectrum is governed by two $N = 2$ supersymmetric quantum-mechanical systems.
The $N = 2$ supersymmetric structure is expected to be a common feature in any theories with local symmetries.

In Section \ref{sec:5dGauge}, we saw that each nonzero mode $A_{\mu}^{(n)}$ ($n\neq0$) becomes massive by absorbing $A_{5}^{(n)}$ into the longitudinal component of $A_{\mu}^{(n)}$.
A one-to-one correspondence between $A_{\mu}^{(n)}$ and $A_{5}^{(n)}$ ($n\neq0$) is ensured by the supersymmetry between the mode functions $f^{(n)}$ and $g^{(n)}$.
Thus, the origin of the supersymmetry lies in the higher-dimensional gauge symmetry.

It is now clear why the 5d gravity  theory possesses {\em two} $N = 2$ supersymmetric systems in the 4d spectrum.
We can take the gauge condition $\phi^{(n)} = 0$ for $n\neq0$ 
by absorbing $\phi^{(n)}$ into the longitudinal component 
of $h_{\mu5}^{(n)}$.
\footnote
{%
Precisely speaking, in the gauge condition $\phi^{(n)} = 0$ for $n\neq0$, $h_{\mu\nu}^{(n)}$ also absorbs $\phi^{(n)}$ through the last term in the first equation of Eq.(\ref{eq:gravity86}).
}
The supersymmetry between the mode functions $g^{(n)}$ and $k^{(n)}$ ensures a one-to-one correspondence between $h_{\mu5}^{(n)}$ and $\phi^{(n)}$.
We can further take the gauge condition $h_{\mu5}^{(n)} = 0$ for $n\neq0$ and then $h_{\mu\nu}^{(n)}$ becomes massive by absorbing $h_{\mu5}^{(n)}$ into $h_{\mu\nu}^{(n)}$.
The supersymmetry between $f^{(n)}$ and $g^{(n)}$ ensures a one-to-one correspondence between $h_{\mu\nu}^{(n)}$ and $h_{\mu5}^{(n)}$.
Thus, the origin of the two $N = 2$ supersymmetric structures lies in the higher-dimensional general coordinate invariance.
It should be emphasized that the full supersymmetry is lost if unphysical degrees of freedom are removed from the spectrum.

We have discussed the boundary conditions from a supersymmetric point of view and succeeded to derive the allowed boundary conditions compatible with the supersymmetry, which are consistent with those obtained from a geometrical point of view \cite{CGR:2000,GRS:2000,PRZ:2000,LM:2001,DL:2003,GPP:2005,CLP:2005,BCLPS:2006}.
It will be interesting to point out the difference of the allowed boundary conditions between the 5d gauge and the 5d gravity theories.
In the 5d gauge theory there are four types of the allowed boundary conditions.
If we choose any boundary conditions other than the type (N, N), the 4d gauge symmetry is broken because no massless 4d vector appears, as depicted in FIG. \ref{fig:spectrum}.
On the other hand, in the 5d gravity theory there is only a unique set of the boundary conditions, for which a massless graviton and a massless scalar appear.

We have shown that the mass eigenvalue $m_{n}$ of the mode functions $f^{(n)}$, $g^{(n)}$ and $k^{(n)}$ is triply degenerate (except for the zero modes).
A pair of $\{f^{(n)}, g^{(n)}\}$ form a supersymmetry multiplet in an $N = 2$ supersymmetric quantum-mechanical system.
Furthermore, a pair of $\{g^{(n)}, k^{(n)}\}$ form a supersymmetry multiplet in another $N = 2$ supersymmetric one.
We then expect that the two $N = 2$ supersymmetric systems would be embedded in a system with some extension of the $N = 2$ supersymmetry.
The extension cannot be, however, the standard $N$-extended one because the number of the degeneracy between ``bosonic'' and ``fermionic'' degrees of freedom is necessarily even.
We have discussed two possible extensions \cite{ACDI:1995,AIN:1995,Fernandez:1997,AIN:1999,AST:2001,CIN:2002,AS:2003} \cite{LNOSS3:2007}.
An interesting fact is that both of them are nonlinear extensions of the $N = 2$ supersymmetry algebra.
It would be of great interest to explore possible extensions of the standard supersymmetry.

Although the quantum-mechanical supersymmetry is shown  to exist in the 4d spectrum, it does not imply that the theory possesses the supersymmetry because we have not shown that the action is indeed invariant under some supersymmetry transformations.
Interestingly, in \cite{LNOSS2:2007}, by taking the $R_{\xi}$-gauge the quadratic action of the 5d gravity theory has been shown to be invariant under transformations that are closely related to the supersymmetry found in this paper.
It is, however, unclear that the invariance preserves in the full action.
Further study should be done.

Our study is far from satisfactory.
It will be worthwhile continuing further investigation on this subject.
We should show whether or not the supersymmetric structure found in this paper exists in any higher-dimensional gravity theories.
More importantly, we should clarify physical roles and importance of the supersymmetry in higher-dimensional gauge/gravity theories.
The work will be reported elsewhere.
\section*{ACKNOWLEDGMENTS}
C.S.L. and M.S. are supported in part by the Grant-in-Aid for
Scientific Research (No.18204024 and No.18540275) 
by the Japanese Ministry
of Education, Science, Sports and Culture.
The authors would like to thank K.Ghoroku, N.Maru, K.Sato, H.Sonoda,
M.Tachibana and K.Takenaga
for valuable discussions.
\newpage


\begin{thebibliography}{70}
\expandafter\ifx\csname natexlab\endcsname\relax\def\natexlab#1{#1}\fi
\expandafter\ifx\csname bibnamefont\endcsname\relax
  \def\bibnamefont#1{#1}\fi
\expandafter\ifx\csname bibfnamefont\endcsname\relax
  \def\bibfnamefont#1{#1}\fi
\expandafter\ifx\csname citenamefont\endcsname\relax
  \def\citenamefont#1{#1}\fi
\expandafter\ifx\csname url\endcsname\relax
  \def\url#1{\texttt{#1}}\fi
\expandafter\ifx\csname urlprefix\endcsname\relax\def\urlprefix{URL }\fi
\providecommand{\bibinfo}[2]{#2}
\providecommand{\eprint}[2][]{\url{#2}}
\bibitem[{\citenamefont{Hosotani}(1983)}]{Hosotani:1983}
\bibinfo{author}{\bibfnamefont{Y.}~\bibnamefont{Hosotani}},
  \bibinfo{journal}{Phys. Lett.} \textbf{\bibinfo{volume}{B126}},
  \bibinfo{pages}{309} (\bibinfo{year}{1983}).
\bibitem[{\citenamefont{Hosotani}(1989)}]{Hosotani:1989}
\bibinfo{author}{\bibfnamefont{Y.}~\bibnamefont{Hosotani}},
  \bibinfo{journal}{Ann. Phys.} \textbf{\bibinfo{volume}{190}},
  \bibinfo{pages}{233} (\bibinfo{year}{1989}).
\bibitem[{\citenamefont{Hatanaka et~al.}(1998)\citenamefont{Hatanaka, Inami,
  and Lim}}]{HIL:1998}
\bibinfo{author}{\bibfnamefont{H.}~\bibnamefont{Hatanaka}},
  \bibinfo{author}{\bibfnamefont{T.}~\bibnamefont{Inami}}, \bibnamefont{and}
  \bibinfo{author}{\bibfnamefont{C.~S.} \bibnamefont{Lim}},
  \bibinfo{journal}{Mod. Phys. Lett.} \textbf{\bibinfo{volume}{A13}},
  \bibinfo{pages}{2601} (\bibinfo{year}{1998}), \eprint{hep-th/9805067}.
\bibitem[{\citenamefont{Antoniadis et~al.}(2001)\citenamefont{Antoniadis,
  Benakli, and Quiros}}]{ABQ:2001}
\bibinfo{author}{\bibfnamefont{I.}~\bibnamefont{Antoniadis}},
  \bibinfo{author}{\bibfnamefont{K.}~\bibnamefont{Benakli}}, \bibnamefont{and}
  \bibinfo{author}{\bibfnamefont{M.}~\bibnamefont{Quiros}},
  \bibinfo{journal}{New J. Phys.} \textbf{\bibinfo{volume}{3}},
  \bibinfo{pages}{20} (\bibinfo{year}{2001}), \eprint{hep-th/0108005}.
\bibitem[{\citenamefont{Kubo et~al.}(2002)\citenamefont{Kubo, Lim, and
  Yamashita}}]{KLY:2002}
\bibinfo{author}{\bibfnamefont{M.}~\bibnamefont{Kubo}},
  \bibinfo{author}{\bibfnamefont{C.~S.} \bibnamefont{Lim}}, \bibnamefont{and}
  \bibinfo{author}{\bibfnamefont{H.}~\bibnamefont{Yamashita}},
  \bibinfo{journal}{Mod. Phys. Lett.} \textbf{\bibinfo{volume}{A17}},
  \bibinfo{pages}{2249} (\bibinfo{year}{2002}), \eprint{hep-ph/0111327}.
\bibitem[{\citenamefont{Hall et~al.}(2002)\citenamefont{Hall, Nomura, and
  Smith}}]{HNS:2002}
\bibinfo{author}{\bibfnamefont{L.~J.} \bibnamefont{Hall}},
  \bibinfo{author}{\bibfnamefont{Y.}~\bibnamefont{Nomura}}, \bibnamefont{and}
  \bibinfo{author}{\bibfnamefont{D.~R.} \bibnamefont{Smith}},
  \bibinfo{journal}{Nucl. Phys.} \textbf{\bibinfo{volume}{B639}},
  \bibinfo{pages}{307} (\bibinfo{year}{2002}), \eprint{hep-ph/0107331}.
\bibitem[{\citenamefont{Burdman and Nomura}(2003)}]{BN:2003}
\bibinfo{author}{\bibfnamefont{G.}~\bibnamefont{Burdman}} \bibnamefont{and}
  \bibinfo{author}{\bibfnamefont{Y.}~\bibnamefont{Nomura}},
  \bibinfo{journal}{Nucl. Phys.} \textbf{\bibinfo{volume}{B656}},
  \bibinfo{pages}{3} (\bibinfo{year}{2003}), \eprint{hep-ph/0210257}.
\bibitem[{\citenamefont{Haba et~al.}(2004)\citenamefont{Haba, Hosotani,
  Kawamura, and Yamashita}}]{HHKY:2004}
\bibinfo{author}{\bibfnamefont{N.}~\bibnamefont{Haba}},
  \bibinfo{author}{\bibfnamefont{Y.}~\bibnamefont{Hosotani}},
  \bibinfo{author}{\bibfnamefont{Y.}~\bibnamefont{Kawamura}}, \bibnamefont{and}
  \bibinfo{author}{\bibfnamefont{T.}~\bibnamefont{Yamashita}},
  \bibinfo{journal}{Phys. Rev.} \textbf{\bibinfo{volume}{D70}},
  \bibinfo{pages}{015010} (\bibinfo{year}{2004}), \eprint{hep-ph/0401183}.
\bibitem[{\citenamefont{Haba et~al.}(2005)\citenamefont{Haba, Takenaga, and
  Yamashita}}]{HTY:2005}
\bibinfo{author}{\bibfnamefont{N.}~\bibnamefont{Haba}},
  \bibinfo{author}{\bibfnamefont{K.}~\bibnamefont{Takenaga}}, \bibnamefont{and}
  \bibinfo{author}{\bibfnamefont{T.}~\bibnamefont{Yamashita}},
  \bibinfo{journal}{Phys. Lett.} \textbf{\bibinfo{volume}{B615}},
  \bibinfo{pages}{247} (\bibinfo{year}{2005}), \eprint{hep-ph/0411250}.
\bibitem[{\citenamefont{Hosotani et~al.}(2004)\citenamefont{Hosotani, Noda, and
  Takenaga}}]{HNT:2004}
\bibinfo{author}{\bibfnamefont{Y.}~\bibnamefont{Hosotani}},
  \bibinfo{author}{\bibfnamefont{S.}~\bibnamefont{Noda}}, \bibnamefont{and}
  \bibinfo{author}{\bibfnamefont{K.}~\bibnamefont{Takenaga}},
  \bibinfo{journal}{Phys. Rev.} \textbf{\bibinfo{volume}{D69}},
  \bibinfo{pages}{125014} (\bibinfo{year}{2004}), \eprint{hep-ph/0403106}.
\bibitem[{\citenamefont{Hosotani et~al.}(2005)\citenamefont{Hosotani, Noda, and
  Takenaga}}]{HNT:2005}
\bibinfo{author}{\bibfnamefont{Y.}~\bibnamefont{Hosotani}},
  \bibinfo{author}{\bibfnamefont{S.}~\bibnamefont{Noda}}, \bibnamefont{and}
  \bibinfo{author}{\bibfnamefont{K.}~\bibnamefont{Takenaga}},
  \bibinfo{journal}{Phys. Lett.} \textbf{\bibinfo{volume}{B607}},
  \bibinfo{pages}{276} (\bibinfo{year}{2005}), \eprint{hep-ph/0410193}.
\bibitem[{\citenamefont{Kawamura}(2001{\natexlab{a}})}]{Kawamura1:2001}
\bibinfo{author}{\bibfnamefont{Y.}~\bibnamefont{Kawamura}},
  \bibinfo{journal}{Prog. Theor. Phys.} \textbf{\bibinfo{volume}{105}},
  \bibinfo{pages}{691} (\bibinfo{year}{2001}{\natexlab{a}}),
  \eprint{hep-ph/0012352}.
\bibitem[{\citenamefont{Kawamura}(2001{\natexlab{b}})}]{Kawamura2:2001}
\bibinfo{author}{\bibfnamefont{Y.}~\bibnamefont{Kawamura}},
  \bibinfo{journal}{Prog. Theor. Phys.} \textbf{\bibinfo{volume}{105}},
  \bibinfo{pages}{999 } (\bibinfo{year}{2001}{\natexlab{b}}),
  \eprint{hep-ph/0012125}.
\bibitem[{\citenamefont{Hall and Nomura}(2001)}]{HN:2001}
\bibinfo{author}{\bibfnamefont{L.}~\bibnamefont{Hall}} \bibnamefont{and}
  \bibinfo{author}{\bibfnamefont{Y.}~\bibnamefont{Nomura}},
  \bibinfo{journal}{Phys. Rev.} \textbf{\bibinfo{volume}{D64}},
  \bibinfo{pages}{055003} (\bibinfo{year}{2001}), \eprint{hep-ph/0103125}.
\bibitem[{\citenamefont{Altarelli and Feruglio}(2001)}]{AF:2001}
\bibinfo{author}{\bibfnamefont{G.}~\bibnamefont{Altarelli}} \bibnamefont{and}
  \bibinfo{author}{\bibfnamefont{F.}~\bibnamefont{Feruglio}},
  \bibinfo{journal}{Phys. Lett.} \textbf{\bibinfo{volume}{B511}},
  \bibinfo{pages}{257 } (\bibinfo{year}{2001}), \eprint{hep-ph/0102301}.
\bibitem[{\citenamefont{Hebecker and March-Russell}(2001)}]{HM-R:2001}
\bibinfo{author}{\bibfnamefont{A.}~\bibnamefont{Hebecker}} \bibnamefont{and}
  \bibinfo{author}{\bibfnamefont{J.}~\bibnamefont{March-Russell}},
  \bibinfo{journal}{Nucl. Phys.} \textbf{\bibinfo{volume}{B613}},
  \bibinfo{pages}{3 } (\bibinfo{year}{2001}), \eprint{hep-ph/0106166}.
\bibitem[{\citenamefont{Nagasawa and Sakamoto}(2004)}]{NS:2004}
\bibinfo{author}{\bibfnamefont{T.}~\bibnamefont{Nagasawa}} \bibnamefont{and}
  \bibinfo{author}{\bibfnamefont{M.}~\bibnamefont{Sakamoto}},
  \bibinfo{journal}{Prog. Theor. Phys.} \textbf{\bibinfo{volume}{112}},
  \bibinfo{pages}{629} (\bibinfo{year}{2004}), \eprint{hep-ph/0406024}.
\bibitem[{\citenamefont{Cs\'aki
  et~al.}(2004{\natexlab{a}})\citenamefont{Cs\'aki, Grojean, Murayama, Pilo,
  and Terning}}]{Csaki:2003dt}
\bibinfo{author}{\bibfnamefont{C.}~\bibnamefont{Cs\'aki}},
  \bibinfo{author}{\bibfnamefont{C.}~\bibnamefont{Grojean}},
  \bibinfo{author}{\bibfnamefont{H.}~\bibnamefont{Murayama}},
  \bibinfo{author}{\bibfnamefont{L.}~\bibnamefont{Pilo}}, \bibnamefont{and}
  \bibinfo{author}{\bibfnamefont{J.}~\bibnamefont{Terning}},
  \bibinfo{journal}{Phys. Rev.} \textbf{\bibinfo{volume}{D69}},
  \bibinfo{pages}{055006} (\bibinfo{year}{2004}{\natexlab{a}}),
  \eprint{hep-ph/0305237}.
\bibitem[{\citenamefont{Chivukula et~al.}(2004)\citenamefont{Chivukula,
  Simmons, He, Kurachi, and Tanabashi}}]{Chivukula:2004pk}
\bibinfo{author}{\bibfnamefont{R.~S.} \bibnamefont{Chivukula}},
  \bibinfo{author}{\bibfnamefont{E.~H.} \bibnamefont{Simmons}},
  \bibinfo{author}{\bibfnamefont{H.-J.} \bibnamefont{He}},
  \bibinfo{author}{\bibfnamefont{M.}~\bibnamefont{Kurachi}}, \bibnamefont{and}
  \bibinfo{author}{\bibfnamefont{M.}~\bibnamefont{Tanabashi}},
  \bibinfo{journal}{Phys. Rev.} \textbf{\bibinfo{volume}{D70}},
  \bibinfo{pages}{075008} (\bibinfo{year}{2004}), \eprint{hep-ph/0406077}.
\bibitem[{\citenamefont{Georgi}(2005)}]{Georgi:2004iy}
\bibinfo{author}{\bibfnamefont{H.}~\bibnamefont{Georgi}},
  \bibinfo{journal}{Phys. Rev.} \textbf{\bibinfo{volume}{D71}},
  \bibinfo{pages}{015016} (\bibinfo{year}{2005}), \eprint{hep-ph/0408067}.
\bibitem[{\citenamefont{Foadi et~al.}(2004)\citenamefont{Foadi, Gopalakrishna,
  and Schmidt}}]{Foadi:2003xa}
\bibinfo{author}{\bibfnamefont{R.}~\bibnamefont{Foadi}},
  \bibinfo{author}{\bibfnamefont{S.}~\bibnamefont{Gopalakrishna}},
  \bibnamefont{and} \bibinfo{author}{\bibfnamefont{C.}~\bibnamefont{Schmidt}},
  \bibinfo{journal}{JHEP} \textbf{\bibinfo{volume}{03}}, \bibinfo{pages}{042}
  (\bibinfo{year}{2004}), \eprint{hep-ph/0312324}.
\bibitem[{\citenamefont{Cs\'aki
  et~al.}(2004{\natexlab{b}})\citenamefont{Cs\'aki, Grojean, Pilo, and
  Terning}}]{Csaki:2003zu}
\bibinfo{author}{\bibfnamefont{C.}~\bibnamefont{Cs\'aki}},
  \bibinfo{author}{\bibfnamefont{C.}~\bibnamefont{Grojean}},
  \bibinfo{author}{\bibfnamefont{L.}~\bibnamefont{Pilo}}, \bibnamefont{and}
  \bibinfo{author}{\bibfnamefont{J.}~\bibnamefont{Terning}},
  \bibinfo{journal}{Phys. Rev. Lett.} \textbf{\bibinfo{volume}{92}},
  \bibinfo{pages}{101802} (\bibinfo{year}{2004}{\natexlab{b}}),
  \eprint{hep-ph/0308038}.
\bibitem[{\citenamefont{Nomura}(2003)}]{Nomura:2003du}
\bibinfo{author}{\bibfnamefont{Y.}~\bibnamefont{Nomura}},
  \bibinfo{journal}{JHEP} \textbf{\bibinfo{volume}{11}}, \bibinfo{pages}{050}
  (\bibinfo{year}{2003}), \eprint{hep-ph/0309189}.
\bibitem[{\citenamefont{Gabriel et~al.}(2004)\citenamefont{Gabriel, Nandi, and
  Seidl}}]{Gabriel:2004ua}
\bibinfo{author}{\bibfnamefont{S.}~\bibnamefont{Gabriel}},
  \bibinfo{author}{\bibfnamefont{S.}~\bibnamefont{Nandi}}, \bibnamefont{and}
  \bibinfo{author}{\bibfnamefont{G.}~\bibnamefont{Seidl}},
  \bibinfo{journal}{Phys. Lett.} \textbf{\bibinfo{volume}{B603}},
  \bibinfo{pages}{74} (\bibinfo{year}{2004}), \eprint{hep-ph/0406020}.
\bibitem[{\citenamefont{Schwinn}(2004)}]{Schwinn:2004xa}
\bibinfo{author}{\bibfnamefont{C.}~\bibnamefont{Schwinn}},
  \bibinfo{journal}{Phys. Rev.} \textbf{\bibinfo{volume}{D69}},
  \bibinfo{pages}{116005} (\bibinfo{year}{2004}), \eprint{hep-ph/0402118}.
\bibitem[{\citenamefont{Cacciapaglia et~al.}(2004)\citenamefont{Cacciapaglia,
  Cs\'aki, Grojean, and Terning}}]{Cacciapaglia:2004jz}
\bibinfo{author}{\bibfnamefont{G.}~\bibnamefont{Cacciapaglia}},
  \bibinfo{author}{\bibfnamefont{C.}~\bibnamefont{Cs\'aki}},
  \bibinfo{author}{\bibfnamefont{C.}~\bibnamefont{Grojean}}, \bibnamefont{and}
  \bibinfo{author}{\bibfnamefont{J.}~\bibnamefont{Terning}},
  \bibinfo{journal}{Phys. Rev.} \textbf{\bibinfo{volume}{D70}},
  \bibinfo{pages}{075014} (\bibinfo{year}{2004}), \eprint{hep-ph/0401160}.
\bibitem[{\citenamefont{Burdman and Nomura}(2004)}]{Burdman:2003ya}
\bibinfo{author}{\bibfnamefont{G.}~\bibnamefont{Burdman}} \bibnamefont{and}
  \bibinfo{author}{\bibfnamefont{Y.}~\bibnamefont{Nomura}},
  \bibinfo{journal}{Phys. Rev.} \textbf{\bibinfo{volume}{D69}},
  \bibinfo{pages}{115013} (\bibinfo{year}{2004}), \eprint{hep-ph/0312247}.
\bibitem[{\citenamefont{Cs\'aki}(2004)}]{Csaki:2004sz}
\bibinfo{author}{\bibfnamefont{C.}~\bibnamefont{Cs\'aki}}
  (\bibinfo{year}{2004}), \eprint{hep-ph/0412339}.
\bibitem[{\citenamefont{Randall and Sundrum}(1999{\natexlab{a}})}]{RS1:1999}
\bibinfo{author}{\bibfnamefont{L.}~\bibnamefont{Randall}} \bibnamefont{and}
  \bibinfo{author}{\bibfnamefont{R.}~\bibnamefont{Sundrum}},
  \bibinfo{journal}{Phys. Rev. Lett.} \textbf{\bibinfo{volume}{83}},
  \bibinfo{pages}{3370} (\bibinfo{year}{1999}{\natexlab{a}}),
  \eprint{hep-ph/9905221}.
\bibitem[{\citenamefont{Goldberger et~al.}(2003)\citenamefont{Goldberger,
  Nomura, and Smith}}]{Goldberger:2002pc}
\bibinfo{author}{\bibfnamefont{W.~D.} \bibnamefont{Goldberger}},
  \bibinfo{author}{\bibfnamefont{Y.}~\bibnamefont{Nomura}}, \bibnamefont{and}
  \bibinfo{author}{\bibfnamefont{D.~R.} \bibnamefont{Smith}},
  \bibinfo{journal}{Phys. Rev.} \textbf{\bibinfo{volume}{D67}},
  \bibinfo{pages}{075021} (\bibinfo{year}{2003}), \eprint{hep-ph/0209158}.
\bibitem[{\citenamefont{Pomarol}(2000)}]{Pomarol:1999ad}
\bibinfo{author}{\bibfnamefont{A.}~\bibnamefont{Pomarol}},
  \bibinfo{journal}{Phys. Lett.} \textbf{\bibinfo{volume}{B486}},
  \bibinfo{pages}{153} (\bibinfo{year}{2000}), \eprint{hep-ph/9911294}.
\bibitem[{\citenamefont{Davoudiasl et~al.}(2000)\citenamefont{Davoudiasl,
  Hewett, and Rizzo}}]{Davoudiasl:1999tf}
\bibinfo{author}{\bibfnamefont{H.}~\bibnamefont{Davoudiasl}},
  \bibinfo{author}{\bibfnamefont{J.~L.} \bibnamefont{Hewett}},
  \bibnamefont{and} \bibinfo{author}{\bibfnamefont{T.~G.} \bibnamefont{Rizzo}},
  \bibinfo{journal}{Phys. Lett.} \textbf{\bibinfo{volume}{B473}},
  \bibinfo{pages}{43} (\bibinfo{year}{2000}), \eprint{hep-ph/9911262}.
\bibitem[{\citenamefont{Randall and Sundrum}(1999{\natexlab{b}})}]{RS2:1999}
\bibinfo{author}{\bibfnamefont{L.}~\bibnamefont{Randall}} \bibnamefont{and}
  \bibinfo{author}{\bibfnamefont{R.}~\bibnamefont{Sundrum}},
  \bibinfo{journal}{Phys. Rev. Lett.} \textbf{\bibinfo{volume}{83}},
  \bibinfo{pages}{4690} (\bibinfo{year}{1999}{\natexlab{b}}),
  \eprint{hep-th/9906064}.
\bibitem[{\citenamefont{Kim and Kim}(2000)}]{Kim:1999ja}
\bibinfo{author}{\bibfnamefont{H.~B.} \bibnamefont{Kim}} \bibnamefont{and}
  \bibinfo{author}{\bibfnamefont{H.~D.} \bibnamefont{Kim}},
  \bibinfo{journal}{Phys. Rev.} \textbf{\bibinfo{volume}{D61}},
  \bibinfo{pages}{064003} (\bibinfo{year}{2000}), \eprint{hep-th/9909053}.
\bibitem[{\citenamefont{Lykken and Randall}(2000)}]{Lykken:1999nb}
\bibinfo{author}{\bibfnamefont{J.~D.} \bibnamefont{Lykken}} \bibnamefont{and}
  \bibinfo{author}{\bibfnamefont{L.}~\bibnamefont{Randall}},
  \bibinfo{journal}{JHEP} \textbf{\bibinfo{volume}{06}}, \bibinfo{pages}{014}
  (\bibinfo{year}{2000}), \eprint{hep-th/9908076}.
\bibitem[{\citenamefont{Goldberger and Wise}(1999)}]{Goldberger:1999uk}
\bibinfo{author}{\bibfnamefont{W.~D.} \bibnamefont{Goldberger}}
  \bibnamefont{and} \bibinfo{author}{\bibfnamefont{M.~B.} \bibnamefont{Wise}},
  \bibinfo{journal}{Phys. Rev. Lett.} \textbf{\bibinfo{volume}{83}},
  \bibinfo{pages}{4922} (\bibinfo{year}{1999}), \eprint{hep-ph/9907447}.
\bibitem[{\citenamefont{Chang et~al.}(2000)\citenamefont{Chang, Hisano, Nakano,
  Okada, and Yamaguchi}}]{Chang:1999nh}
\bibinfo{author}{\bibfnamefont{S.}~\bibnamefont{Chang}},
  \bibinfo{author}{\bibfnamefont{J.}~\bibnamefont{Hisano}},
  \bibinfo{author}{\bibfnamefont{H.}~\bibnamefont{Nakano}},
  \bibinfo{author}{\bibfnamefont{N.}~\bibnamefont{Okada}}, \bibnamefont{and}
  \bibinfo{author}{\bibfnamefont{M.}~\bibnamefont{Yamaguchi}},
  \bibinfo{journal}{Phys. Rev.} \textbf{\bibinfo{volume}{D62}},
  \bibinfo{pages}{084025} (\bibinfo{year}{2000}), \eprint{hep-ph/9912498}.
\bibitem[{\citenamefont{DeWolfe et~al.}(2000)\citenamefont{DeWolfe, Freedman,
  Gubser, and Karch}}]{DeWolfe:1999cp}
\bibinfo{author}{\bibfnamefont{O.}~\bibnamefont{DeWolfe}},
  \bibinfo{author}{\bibfnamefont{D.~Z.} \bibnamefont{Freedman}},
  \bibinfo{author}{\bibfnamefont{S.~S.} \bibnamefont{Gubser}},
  \bibnamefont{and} \bibinfo{author}{\bibfnamefont{A.}~\bibnamefont{Karch}},
  \bibinfo{journal}{Phys. Rev.} \textbf{\bibinfo{volume}{D62}},
  \bibinfo{pages}{046008} (\bibinfo{year}{2000}), \eprint{hep-th/9909134}.
\bibitem[{\citenamefont{Charmousis
  et~al.}(2000{\natexlab{a}})\citenamefont{Charmousis, Gregory, and
  Rubakov}}]{Charmousis:1999rg}
\bibinfo{author}{\bibfnamefont{C.}~\bibnamefont{Charmousis}},
  \bibinfo{author}{\bibfnamefont{R.}~\bibnamefont{Gregory}}, \bibnamefont{and}
  \bibinfo{author}{\bibfnamefont{V.~A.} \bibnamefont{Rubakov}},
  \bibinfo{journal}{Phys. Rev.} \textbf{\bibinfo{volume}{D62}},
  \bibinfo{pages}{067505} (\bibinfo{year}{2000}{\natexlab{a}}),
  \eprint{hep-th/9912160}.
\bibitem[{\citenamefont{Cs\'aki et~al.}(1999)\citenamefont{Cs\'aki, Graesser,
  Kolda, and Terning}}]{Csaki:1999jh}
\bibinfo{author}{\bibfnamefont{C.}~\bibnamefont{Cs\'aki}},
  \bibinfo{author}{\bibfnamefont{M.}~\bibnamefont{Graesser}},
  \bibinfo{author}{\bibfnamefont{C.~F.} \bibnamefont{Kolda}}, \bibnamefont{and}
  \bibinfo{author}{\bibfnamefont{J.}~\bibnamefont{Terning}},
  \bibinfo{journal}{Phys. Lett.} \textbf{\bibinfo{volume}{B462}},
  \bibinfo{pages}{34} (\bibinfo{year}{1999}), \eprint{hep-ph/9906513}.
\bibitem[{\citenamefont{Maru and Yamashita}(2006)}]{MY:2006}
\bibinfo{author}{\bibfnamefont{N.}~\bibnamefont{Maru}} \bibnamefont{and}
  \bibinfo{author}{\bibfnamefont{T.}~\bibnamefont{Yamashita}},
  \bibinfo{journal}{Nucl. Phys.} \textbf{\bibinfo{volume}{B754}},
  \bibinfo{pages}{127} (\bibinfo{year}{2006}), \eprint{hep-ph/0603237}.
\bibitem[{\citenamefont{Hosotani}(2006)}]{Hosotani:2006}
\bibinfo{author}{\bibfnamefont{Y.}~\bibnamefont{Hosotani}}
  (\bibinfo{year}{2006}), \eprint{hep-ph/0607064}.
\bibitem[{\citenamefont{Hosotani et~al.}(2007)\citenamefont{Hosotani, Maru,
  Takenaga, and Yamashita}}]{Hosotani:2007kn}
\bibinfo{author}{\bibfnamefont{Y.}~\bibnamefont{Hosotani}},
  \bibinfo{author}{\bibfnamefont{N.}~\bibnamefont{Maru}},
  \bibinfo{author}{\bibfnamefont{K.}~\bibnamefont{Takenaga}}, \bibnamefont{and}
  \bibinfo{author}{\bibfnamefont{T.}~\bibnamefont{Yamashita}}
  (\bibinfo{year}{2007}), \eprint{0709.2844 [hep-ph]}.
\bibitem[{\citenamefont{Lim et~al.}(2005)\citenamefont{Lim, Nagasawa, Sakamoto,
  and Sonoda}}]{LNSS:2005}
\bibinfo{author}{\bibfnamefont{C.~S.} \bibnamefont{Lim}},
  \bibinfo{author}{\bibfnamefont{T.}~\bibnamefont{Nagasawa}},
  \bibinfo{author}{\bibfnamefont{M.}~\bibnamefont{Sakamoto}}, \bibnamefont{and}
  \bibinfo{author}{\bibfnamefont{H.}~\bibnamefont{Sonoda}},
  \bibinfo{journal}{Phys. Rev.} \textbf{\bibinfo{volume}{D72}},
  \bibinfo{pages}{064006} (\bibinfo{year}{2005}), \eprint{hep-th/0502022}.
\bibitem[{\citenamefont{Howe et~al.}(1989)\citenamefont{Howe, Penati, Pernici,
  and Townsend}}]{HPPT:1989}
\bibinfo{author}{\bibfnamefont{P.~S.} \bibnamefont{Howe}},
  \bibinfo{author}{\bibfnamefont{S.}~\bibnamefont{Penati}},
  \bibinfo{author}{\bibfnamefont{M.}~\bibnamefont{Pernici}}, \bibnamefont{and}
  \bibinfo{author}{\bibfnamefont{P.~K.} \bibnamefont{Townsend}},
  \bibinfo{journal}{Class. Quant. Grav.} \textbf{\bibinfo{volume}{6}},
  \bibinfo{pages}{1125} (\bibinfo{year}{1989}).
\bibitem[{\citenamefont{Witten}(1981)}]{Witten:1981}
\bibinfo{author}{\bibfnamefont{E.}~\bibnamefont{Witten}},
  \bibinfo{journal}{Nucl. Phys.} \textbf{\bibinfo{volume}{B188}},
  \bibinfo{pages}{513} (\bibinfo{year}{1981}).
\bibitem[{\citenamefont{Miemiec}(2001)}]{Miemiec:2001}
\bibinfo{author}{\bibfnamefont{A.}~\bibnamefont{Miemiec}},
  \bibinfo{journal}{Fortsch. Phys.} \textbf{\bibinfo{volume}{49}},
  \bibinfo{pages}{747} (\bibinfo{year}{2001}), \eprint{hep-th/0011160}.
\bibitem[{\citenamefont{Nagasawa et~al.}(2003)\citenamefont{Nagasawa, Sakamoto,
  and Takenaga}}]{NST:2003}
\bibinfo{author}{\bibfnamefont{T.}~\bibnamefont{Nagasawa}},
  \bibinfo{author}{\bibfnamefont{M.}~\bibnamefont{Sakamoto}}, \bibnamefont{and}
  \bibinfo{author}{\bibfnamefont{K.}~\bibnamefont{Takenaga}},
  \bibinfo{journal}{Phys. Lett.} \textbf{\bibinfo{volume}{B562}},
  \bibinfo{pages}{358} (\bibinfo{year}{2003}), \eprint{hep-th/0212192}.
\bibitem[{\citenamefont{Nagasawa et~al.}(2004)\citenamefont{Nagasawa, Sakamoto,
  and Takenaga}}]{NST:2004}
\bibinfo{author}{\bibfnamefont{T.}~\bibnamefont{Nagasawa}},
  \bibinfo{author}{\bibfnamefont{M.}~\bibnamefont{Sakamoto}}, \bibnamefont{and}
  \bibinfo{author}{\bibfnamefont{K.}~\bibnamefont{Takenaga}},
  \bibinfo{journal}{Phys. Lett.} \textbf{\bibinfo{volume}{B583}},
  \bibinfo{pages}{357} (\bibinfo{year}{2004}), \eprint{hep-th/0311043}.
\bibitem[{\citenamefont{Nagasawa et~al.}(2005)\citenamefont{Nagasawa, Sakamoto,
  and Takenaga}}]{NST:2005}
\bibinfo{author}{\bibfnamefont{T.}~\bibnamefont{Nagasawa}},
  \bibinfo{author}{\bibfnamefont{M.}~\bibnamefont{Sakamoto}}, \bibnamefont{and}
  \bibinfo{author}{\bibfnamefont{K.}~\bibnamefont{Takenaga}},
  \bibinfo{journal}{J. Phys.} \textbf{\bibinfo{volume}{A38}},
  \bibinfo{pages}{8053} (\bibinfo{year}{2005}), \eprint{hep-th/0505132}.
\bibitem[{\citenamefont{Gherghetta et~al.}(2005)\citenamefont{Gherghetta,
  Peloso, and Poppitz}}]{GPP:2005}
\bibinfo{author}{\bibfnamefont{T.}~\bibnamefont{Gherghetta}},
  \bibinfo{author}{\bibfnamefont{M.}~\bibnamefont{Peloso}}, \bibnamefont{and}
  \bibinfo{author}{\bibfnamefont{E.}~\bibnamefont{Poppitz}},
  \bibinfo{journal}{Phys. Rev.} \textbf{\bibinfo{volume}{D72}},
  \bibinfo{pages}{104003} (\bibinfo{year}{2005}), \eprint{hep-th/0507245}.
\bibitem[{\citenamefont{Chacko et~al.}(2004)\citenamefont{Chacko, Graesser,
  Grojean, and Pilo}}]{CGGP:2004}
\bibinfo{author}{\bibfnamefont{Z.}~\bibnamefont{Chacko}},
  \bibinfo{author}{\bibfnamefont{M.~L.} \bibnamefont{Graesser}},
  \bibinfo{author}{\bibfnamefont{C.}~\bibnamefont{Grojean}}, \bibnamefont{and}
  \bibinfo{author}{\bibfnamefont{L.}~\bibnamefont{Pilo}},
  \bibinfo{journal}{Phys. Rev.} \textbf{\bibinfo{volume}{D70}},
  \bibinfo{pages}{084028} (\bibinfo{year}{2004}), \eprint{hep-th/0312117}.
\bibitem[{\citenamefont{Andrianov
  et~al.}(1995{\natexlab{a}})\citenamefont{Andrianov, Cannata, Dedonder, and
  Ioffe}}]{ACDI:1995}
\bibinfo{author}{\bibfnamefont{A.~A.} \bibnamefont{Andrianov}},
  \bibinfo{author}{\bibfnamefont{F.}~\bibnamefont{Cannata}},
  \bibinfo{author}{\bibfnamefont{J.-P.} \bibnamefont{Dedonder}},
  \bibnamefont{and} \bibinfo{author}{\bibfnamefont{M.~V.} \bibnamefont{Ioffe}},
  \bibinfo{journal}{Int. J. Mod. Phys.} \textbf{\bibinfo{volume}{A10}},
  \bibinfo{pages}{2683} (\bibinfo{year}{1995}{\natexlab{a}}),
  \eprint{hep-th/9404061}.
\bibitem[{\citenamefont{Andrianov
  et~al.}(1995{\natexlab{b}})\citenamefont{Andrianov, Ioffe, and
  Nishnianidze}}]{AIN:1995}
\bibinfo{author}{\bibfnamefont{A.~A.} \bibnamefont{Andrianov}},
  \bibinfo{author}{\bibfnamefont{M.~V.} \bibnamefont{Ioffe}}, \bibnamefont{and}
  \bibinfo{author}{\bibfnamefont{D.~N.} \bibnamefont{Nishnianidze}},
  \bibinfo{journal}{Theor. Math. Phys.} \textbf{\bibinfo{volume}{104}},
  \bibinfo{pages}{1129} (\bibinfo{year}{1995}{\natexlab{b}}).
\bibitem[{\citenamefont{Fernandez~C}(1997)}]{Fernandez:1997}
\bibinfo{author}{\bibfnamefont{D.~J.} \bibnamefont{Fernandez~C}},
  \bibinfo{journal}{Int. J. Mod. Phys.} \textbf{\bibinfo{volume}{A12}},
  \bibinfo{pages}{171} (\bibinfo{year}{1997}), \eprint{quant-ph/9609009}.
\bibitem[{\citenamefont{Andrianov et~al.}(1999)\citenamefont{Andrianov, Ioffe,
  and Nishnianidze}}]{AIN:1999}
\bibinfo{author}{\bibfnamefont{A.~A.} \bibnamefont{Andrianov}},
  \bibinfo{author}{\bibfnamefont{M.~V.} \bibnamefont{Ioffe}}, \bibnamefont{and}
  \bibinfo{author}{\bibfnamefont{D.~N.} \bibnamefont{Nishnianidze}},
  \bibinfo{journal}{J. Phys.} \textbf{\bibinfo{volume}{A32}},
  \bibinfo{pages}{4641} (\bibinfo{year}{1999}), \eprint{solv-int/9810006}.
\bibitem[{\citenamefont{Aoyama et~al.}(2001)\citenamefont{Aoyama, Sato, and
  Tanaka}}]{AST:2001}
\bibinfo{author}{\bibfnamefont{H.}~\bibnamefont{Aoyama}},
  \bibinfo{author}{\bibfnamefont{M.}~\bibnamefont{Sato}}, \bibnamefont{and}
  \bibinfo{author}{\bibfnamefont{T.}~\bibnamefont{Tanaka}},
  \bibinfo{journal}{Nucl. Phys.} \textbf{\bibinfo{volume}{B619}},
  \bibinfo{pages}{105} (\bibinfo{year}{2001}), \eprint{quant-ph/0106037}.
\bibitem[{\citenamefont{Cannata et~al.}(2002)\citenamefont{Cannata, Ioffe, and
  Nishnianidze}}]{CIN:2002}
\bibinfo{author}{\bibfnamefont{F.}~\bibnamefont{Cannata}},
  \bibinfo{author}{\bibfnamefont{M.~V.} \bibnamefont{Ioffe}}, \bibnamefont{and}
  \bibinfo{author}{\bibfnamefont{D.~N.} \bibnamefont{Nishnianidze}},
  \bibinfo{journal}{J. Phys.} \textbf{\bibinfo{volume}{A35}},
  \bibinfo{pages}{1389} (\bibinfo{year}{2002}), \eprint{hep-th/0201080}.
\bibitem[{\citenamefont{Andrianov and Sokolov}(2003)}]{AS:2003}
\bibinfo{author}{\bibfnamefont{A.~A.} \bibnamefont{Andrianov}}
  \bibnamefont{and} \bibinfo{author}{\bibfnamefont{A.~V.}
  \bibnamefont{Sokolov}}, \bibinfo{journal}{Nucl. Phys.}
  \textbf{\bibinfo{volume}{B660}}, \bibinfo{pages}{25} (\bibinfo{year}{2003}),
  \eprint{hep-th/0301062}.
\bibitem[{\citenamefont{Lim et~al.}({\natexlab{a}})\citenamefont{Lim, Nagasawa,
  Ohya, Sakamoto, and Sakamoto}}]{LNOSS3:2007}
\bibinfo{author}{\bibfnamefont{C.~S.} \bibnamefont{Lim}},
  \bibinfo{author}{\bibfnamefont{T.}~\bibnamefont{Nagasawa}},
  \bibinfo{author}{\bibfnamefont{S.}~\bibnamefont{Ohya}},
  \bibinfo{author}{\bibfnamefont{K.}~\bibnamefont{Sakamoto}}, \bibnamefont{and}
  \bibinfo{author}{\bibfnamefont{M.}~\bibnamefont{Sakamoto}}, \bibinfo{note}{to
  appear}.
\bibitem[{\citenamefont{Charmousis
  et~al.}(2000{\natexlab{b}})\citenamefont{Charmousis, Gregory, and
  Rubakov}}]{CGR:2000}
\bibinfo{author}{\bibfnamefont{C.}~\bibnamefont{Charmousis}},
  \bibinfo{author}{\bibfnamefont{R.}~\bibnamefont{Gregory}}, \bibnamefont{and}
  \bibinfo{author}{\bibfnamefont{V.~A.} \bibnamefont{Rubakov}},
  \bibinfo{journal}{Phys. Rev.} \textbf{\bibinfo{volume}{D62}},
  \bibinfo{pages}{067505} (\bibinfo{year}{2000}{\natexlab{b}}),
  \eprint{hep-th/9912160}.
\bibitem[{\citenamefont{Gregory et~al.}(2000)\citenamefont{Gregory, Rubakov,
  and Sibiryakov}}]{GRS:2000}
\bibinfo{author}{\bibfnamefont{R.}~\bibnamefont{Gregory}},
  \bibinfo{author}{\bibfnamefont{V.~A.} \bibnamefont{Rubakov}},
  \bibnamefont{and} \bibinfo{author}{\bibfnamefont{S.~M.}
  \bibnamefont{Sibiryakov}}, \bibinfo{journal}{Phys. Rev. Lett.}
  \textbf{\bibinfo{volume}{84}}, \bibinfo{pages}{5928} (\bibinfo{year}{2000}),
  \eprint{hep-th/0002072}.
\bibitem[{\citenamefont{Pilo et~al.}(2000)\citenamefont{Pilo, Rattazzi, and
  Zaffaroni}}]{PRZ:2000}
\bibinfo{author}{\bibfnamefont{L.}~\bibnamefont{Pilo}},
  \bibinfo{author}{\bibfnamefont{R.}~\bibnamefont{Rattazzi}}, \bibnamefont{and}
  \bibinfo{author}{\bibfnamefont{A.}~\bibnamefont{Zaffaroni}},
  \bibinfo{journal}{JHEP} \textbf{\bibinfo{volume}{07}}, \bibinfo{pages}{056}
  (\bibinfo{year}{2000}), \eprint{hep-th/0004028}.
\bibitem[{\citenamefont{Dubovsky and Libanov}(2003)}]{DL:2003}
\bibinfo{author}{\bibfnamefont{S.~L.} \bibnamefont{Dubovsky}} \bibnamefont{and}
  \bibinfo{author}{\bibfnamefont{M.~V.} \bibnamefont{Libanov}},
  \bibinfo{journal}{JHEP} \textbf{\bibinfo{volume}{11}}, \bibinfo{pages}{038}
  (\bibinfo{year}{2003}), \eprint{hep-th/0309131}.
\bibitem[{\citenamefont{Israel}(1966)}]{Israel:1966}
\bibinfo{author}{\bibfnamefont{W.}~\bibnamefont{Israel}},
  \bibinfo{journal}{Nuovo Cim.} \textbf{\bibinfo{volume}{B44S10}},
  \bibinfo{pages}{1} (\bibinfo{year}{1966}), \bibinfo{note}{errata: Nuovo Cim.
  {\bf B48}, 463 (1967)}.
\bibitem[{\citenamefont{Gibbons and Hawking}(1977)}]{GH:1977}
\bibinfo{author}{\bibfnamefont{G.~W.} \bibnamefont{Gibbons}} \bibnamefont{and}
  \bibinfo{author}{\bibfnamefont{S.~W.} \bibnamefont{Hawking}},
  \bibinfo{journal}{Phys. Rev.} \textbf{\bibinfo{volume}{D15}},
  \bibinfo{pages}{2752} (\bibinfo{year}{1977}).
\bibitem[{\citenamefont{Lalak and Matyszkiewicz}(2001)}]{LM:2001}
\bibinfo{author}{\bibfnamefont{Z.}~\bibnamefont{Lalak}} \bibnamefont{and}
  \bibinfo{author}{\bibfnamefont{R.}~\bibnamefont{Matyszkiewicz}},
  \bibinfo{journal}{JHEP} \textbf{\bibinfo{volume}{11}}, \bibinfo{pages}{027}
  (\bibinfo{year}{2001}), \eprint{hep-th/0110141}.
\bibitem[{\citenamefont{Carena et~al.}(2005)\citenamefont{Carena, Lykken, and
  Park}}]{CLP:2005}
\bibinfo{author}{\bibfnamefont{M.}~\bibnamefont{Carena}},
  \bibinfo{author}{\bibfnamefont{J.}~\bibnamefont{Lykken}}, \bibnamefont{and}
  \bibinfo{author}{\bibfnamefont{M.}~\bibnamefont{Park}},
  \bibinfo{journal}{Phys. Rev.} \textbf{\bibinfo{volume}{D72}},
  \bibinfo{pages}{084017} (\bibinfo{year}{2005}), \eprint{hep-ph/0506305}.
\bibitem[{\citenamefont{Bao et~al.}(2006)\citenamefont{Bao, Carena, Lykken,
  Park, and Santiago}}]{BCLPS:2006}
\bibinfo{author}{\bibfnamefont{R.}~\bibnamefont{Bao}},
  \bibinfo{author}{\bibfnamefont{M.}~\bibnamefont{Carena}},
  \bibinfo{author}{\bibfnamefont{J.}~\bibnamefont{Lykken}},
  \bibinfo{author}{\bibfnamefont{M.}~\bibnamefont{Park}}, \bibnamefont{and}
  \bibinfo{author}{\bibfnamefont{J.}~\bibnamefont{Santiago}},
  \bibinfo{journal}{Phys. Rev.} \textbf{\bibinfo{volume}{D73}},
  \bibinfo{pages}{064026} (\bibinfo{year}{2006}), \eprint{hep-th/0511266}.
\bibitem[{\citenamefont{Lim et~al.}({\natexlab{b}})\citenamefont{Lim, Nagasawa,
  Ohya, Sakamoto, and Sakamoto}}]{LNOSS2:2007}
\bibinfo{author}{\bibfnamefont{C.~S.} \bibnamefont{Lim}},
  \bibinfo{author}{\bibfnamefont{T.}~\bibnamefont{Nagasawa}},
  \bibinfo{author}{\bibfnamefont{S.}~\bibnamefont{Ohya}},
  \bibinfo{author}{\bibfnamefont{K.}~\bibnamefont{Sakamoto}}, \bibnamefont{and}
  \bibinfo{author}{\bibfnamefont{M.}~\bibnamefont{Sakamoto}}, \bibinfo{note}{to
  appear}.
\end{thebibliography}
\end{document}